\documentclass[preprint,showpacs,preprintnumbers,amsmath,amssymb,prc]{revtex4}

\usepackage{epsfig}
\usepackage{graphicx}
\usepackage{dcolumn}
\usepackage{bm}

\def\v#1{\mbox{\boldmath$#1$}}
\def\ket#1{|#1 \rangle}
\def\bra#1{\langle #1|}
\begin{document}

\title{On the role of ground state correlations
in hypernuclear non--mesonic weak decay}

\author{E. Bauer$^{1,2}$}
\email{bauer@fisica.unlp.edu.ar}
\author{G. Garbarino$^3$}

\affiliation{
$^1$Departamento de F\'{\i}sica, Universidad Nacional de La Plata,
C. C. 67, 1900 La Plata, Argentina \\
$^2$Instituto de F\'{\i}sica La Plata,
CONICET, 1900 La Plata, Argentina \\
$^3$Dipartimento di Fisica Teorica, Universit\`a di Torino,
I-10125 Torino, Italy}

\date{\today}

\begin{abstract}
The contribution of ground state correlations (GSC) to the
non--mesonic weak decay of $^{12}_\Lambda$C
and other medium to heavy hypernuclei
is studied within a nuclear matter formalism implemented
in a local density approximation. We adopt
a weak transition potential including the exchange of the complete
octets of pseudoscalar and vector mesons as well as a
residual strong interaction modeled on the Bonn potential.
Leading GSC contributions, at first order in the residual strong interaction,
are introduced on the same footing for all isospin channels
of one-- and two--nucleon induced decays.
Together with fermion antisymmetrization, GSC
turn out to be important for an accurate determination of the decay widths.
Besides opening the two--nucleon stimulated decay channels,
for $^{12}_\Lambda$C GSC
are responsible for 14\% of the rate $\Gamma_1$ while increasing
the $\Gamma_{n}/\Gamma_{p}$ ratio by 4\%.
Our final results for $^{12}_\Lambda$C are: $\Gamma_{\rm NM}=0.98$,
$\Gamma_{n}/\Gamma_{p}=0.34$
and $\Gamma_2/\Gamma_{\rm NM}=0.26$. The saturation property
of $\Gamma_{\rm NM}$ with increasing hypernuclear
mass number is clearly observed. The agreement with data of our
predictions for $\Gamma_{\rm NM}$, $\Gamma_n/\Gamma_p$ and $\Gamma_2$ is rather good.
\end{abstract}

\maketitle

\newpage
\section{Introduction}
\label{intro}
The study of nuclear systems with strangeness
is a relevant question in modern nuclear and hadronic physics \cite{snp},
which also implies important links with astrophysical processes and observables
as well as with QCD, the underlying theory of strong interactions.
Various strange nuclear systems can be studied in the laboratory,
ranging from hypernuclei and kaonic nuclei
to exotic hadronic states such as strangelets, $H$--dibaryons
and pentaquark baryons.
Strangeness production can also be investigated in
relativistic heavy--ion collision experiments,
whose main aim is to establish the existence of a quark--gluon plasma.
Moreover, the cold and dense  matter contained in neutron stars is
expected to be composed by strange hadrons, in the form of hyperons and
Bose--Einstein condensates of kaons, and eventually
by strange quark matter for sufficiently dense systems.

The existence of hypernuclei ---bound systems of non--strange and
strange baryons--- opens up the possibility to study
the hyperon--nucleon and hyperon--hyperon interactions in
both the strong and weak sectors. In turn, such
interactions are important inputs, for instance, when investigating
the macroscopic properties (masses and radii) of neutron stars.
The best studied hypernuclei contain a single $\Lambda$--hyperon.
In a nucleus the $\Lambda$ can decay by emitting
a nucleon and a pion (mesonic mode) as it happens in free space, but
its (weak) interaction with the nucleons opens new channels which are
indicated as non--mesonic decay modes
(for recent reviews see Refs.~\cite{ra98,al02,Ch08,Pa07,Ou05}).
These are the dominant decay
channels of medium--heavy nuclei, where, on the contrary, the mesonic decay
is disfavoured by the Pauli blocking effect on the outgoing nucleon.
In particular, one can distinguish between one-- and two--body induced
decays, $\Lambda N\to nN$ and $\Lambda NN\to nNN$.
The hypernuclear lifetime is given
in terms of the mesonic ($\Gamma_{\rm M}=\Gamma_{\pi^-}+\Gamma_{\pi^0}$)
and non--mesonic decay widths ($\Gamma_{\rm NM}=\Gamma_1+\Gamma_2$)
by $\tau=\hbar/\Gamma_{\rm T}=  \hbar/[\Gamma_{\rm M}+\Gamma_{\rm NM}]$.
The various isospin channels contribute to the
one-- and two--nucleon induced non--mesonic rates as follows:
$\Gamma_1=\Gamma_n+\Gamma_p\equiv \Gamma(\Lambda n\to nn)+\Gamma(\Lambda p\to np)$ and
$\Gamma_2=\Gamma_{nn}+\Gamma_{np}+\Gamma_{pp}\equiv
\Gamma(\Lambda nn\to nnn)+\Gamma(\Lambda np\to nnp)+\Gamma(\Lambda pp\to npp)$.

One should note that, strictly speaking, the only observables
in hypernuclear weak decay are the
lifetime $\tau$, the mesonic rates $\Gamma_{\pi^-}$ and $\Gamma_{\pi^0}$
 and the spectra of the emitted particles (nucleons,
pions and photons). None of the above non--mesonic partial decay rates
($\Gamma_n$, $\Gamma_p$, $\Gamma_{np}$, etc) is an observable from a
quantum--mechanical point of view.
Each one of the possible elementary non--mesonic decays occurs in
the nuclear environment, thus subsequent final state interactions (FSI)
modify the quantum numbers of the weak decay nucleons and
new, secondary nucleons are emitted as well: this prevents
the measurement of any of the non--mesonic partial decay rates.
Instead, the total width $\Gamma_{\rm T}$ can be measured:
being an inclusive quantity, for such a measurement one
has to detect any of the possible products of either mesonic or
non--mesonic decays (typically protons from non--mesonic decays).
The fact that the detected particles
undergo FSI does not appreciably alters the lifetime measurement,
since strong interactions proceeds on a much shorter time scale
than weak decays,
and $\tau^{\rm measured}=\tau+\tau^{\rm strong}\simeq \tau\equiv
\hbar/\Gamma_{\rm T}$.

In order to achieve a proper knowledge of the various decay
mechanisms (in particular of the strangeness--changing baryon--baryon
interactions), a meaningful comparison between theory and experiment
must be possible. The above discussion shows that such a comparison
requires the introduction of non--standard theoretical definitions
for the non--mesonic partial decay rates (which, as mentioned,
are not quantum--mechanical observables)
together with the corresponding experimental methods for determining these
rates. In our opinion, this point has not been adequately addressed in
previous works and, among others, it has impacted on the well--known puzzle
on the ratio $\Gamma_n/\Gamma_p$ between the neutron-- and the
proton--induced non--mesonic rates.

In order to explain how the total non--mesonic rate can be determined in
an experiment, we have to discuss first the measurement of the mesonic rates.
The pion and nucleon emitted in a mesonic decay both have
a momentum of about $100\, {\rm MeV/c}$. Nucleons of a few MeV kinetic energy
cannot be observed as they are below the experimental detection thresholds.
Mesonic decays are thus identified by measuring pions
($\pi^-$'s or $\pi^0\to \gamma \gamma$ decays).
The mesonic width $\Gamma_{\pi^-}$ ($\Gamma_{\pi^0}$) is
determined from the observed $\pi^-$ ($\pi^0\to \gamma \gamma$) energy spectra
and the total width $\Gamma_{\rm T}$. For instance,
$\Gamma^{\rm exp}_{\pi^-}=(N_{\pi^-}/N_{\rm hyp})\Gamma^{\rm exp}_{\rm T}$,
$N_{\pi^-}$ being the total number of detected $\pi^-$'s
and $N_{\rm hyp}$ the total number of produced hypernuclei.
Both these numbers are corrected for the detection efficiencies and
the detector acceptances implied in the measurements.
The mesonic rates measured in this way thus include the effect of
in--medium pion renormalization.
Theoretical models \cite{mesonic-th,gal09} also taking into account
distorted pion waves obtained mesonic widths
in agreement with the experimental values
(in particular, the importance of the pion wave--function distortion
was first demonstrated in the works of Ref.~\cite{mesonic-th}).

The experimental total non--mesonic rate is then obtained
as the difference between the total and the mesonic rates,
$\Gamma^{\rm exp}_{\rm NM}=\Gamma^{\rm exp}_{\rm T}-\Gamma^{\rm exp}_{\rm M}$.
The experimental determination of $\Gamma_{n}/\Gamma_{p}$
is much more involved. Indeed, this ratio must be extracted
from the nucleon emission spectra, and this
requires some theoretical input \cite{ga03,ga04}. FSI are very
important for the non--mesonic processes and nucleons which have
or have not suffered FSI are indistinguishable between each other.
A theoretical simulation of nucleon FSI is thus needed and,
in principle, a coherent sum of both kinds of nucleons must be considered
when evaluating the spectra.
Generally, FSI are accounted for by an intranuclear cascade
model \cite{ra97}, which is a semi--classical scheme.

In the present work we study the non--mesonic weak decay of
hypernuclei ranging from $^{11}_\Lambda$B to $^{208}_\Lambda$Pb
by using a nuclear matter approach
implemented in a local density approximation.
All the possible isospin channels for one-- and two--body induced
mechanisms are included in a
microscopic approach based on the evaluation of Goldstone diagrams.
The partial decay rates are derived by starting from a
two--body weak transition potential.
In particular, we investigate the effect of
ground state correlations (GSC), i.e., the contribution of
nucleon--nucleon correlations in the hypernucleus ground state.
Leading order GSC contributions
will be introduced on the same ground for one-- and two--nucleon
induced processes for the first time.
The general formalism we adopt was established in Refs.~\cite{ba03,ba04}.
The weak transition potential
for the nucleon--nucleon strong interaction contributing to the GSC
we adopt a Bonn potential with the exchange
of $\pi$, $\rho$, $\sigma$ and $\omega$ mesons.

The paper is organized as follows. In Section~\ref{pref} we
start with general considerations about FSI, the definitions
we employ for the weak decay rates as well as
the method usually employed for the determination
of $\Gamma_n/\Gamma_p$ from data on nucleon spectra.
In Section~\ref{miss} we present and discuss the general framework for the
evaluation of the one-- and two--nucleon induced decay widths with the
inclusion of GSC. In Section~\ref{fsi} we make
some further considerations about the evaluation of
the widths and we discuss some former work on the subject.
Explicit expressions for the considered GSC diagrams contributing to
the one--nucleon induced rates are given in Section~\ref{gngpgsc}
and in Appendix~A.
Then, in Section~\ref{results} we present our results and
finally in Section~\ref{conclusions} some conclusions are given.

\section{Preliminary considerations on FSI effects and on the determination of
the weak decay rates}
\label{pref}
The $\Gamma_{n}/\Gamma_{p}$ ratio is defined as the
ratio between the total number of primary (i.e., weak decay) neutron--neutron and
neutron--proton pairs, $N^{\rm wd}_{nn}$ and $N^{\rm wd}_{np}$, emerging
from the processes $\Lambda n\to nn$ and $\Lambda p\to np$, respectively.
Due to nucleon final state interactions and two--body induced decays, the following
inequality is expected for the observables $nn$ and $np$ coincidence numbers,
$N_{nn}$ and $N_{np}$
\cite{ga03}~\footnote{Note that an analogous inequality exists between $\Gamma_n/\Gamma_p$
and the ratio between the total number of emitted neutrons and protons,
$N_{n}/N_{p}$~\cite{ga04}. For the present discussion any of these
two expressions is suitable.}:
\begin{equation}
\label{defcos}
\frac{\Gamma_{n}}{\Gamma_{p}} \equiv \frac{N^{\rm wd}_{nn}}{N^{\rm wd}_{np}}
\neq \frac{N_{nn}}{N_{np}}~.
\end{equation}
Only $N_{nn}/N_{np}$ is a quantum--mechanical observable:
generally, its measurement is affected by thresholds on the
nucleon energy and the pair opening angle \cite{Ou05,KEK,KEK2}.
Theoretical models are thus required to determine
the ``experimental'' value of $\Gamma_n/\Gamma_p$ from
a measurement of $N_{nn}/N_{np}$.
This unusual procedure to determine $(\Gamma_n/\Gamma_p)^{\rm exp}$
makes complete sense provided different models are at disposal and
lead to the same extracted ratio: only in such a case one is allowed to
define this value as the experimental result for $\Gamma_n/\Gamma_p$.
It is thus important to explore the predictions of alternative models
when applied to the analysis of data.
In the present section we go deeper into questions of this kind to show
some ambiguities which need to be emphasized for a meaningful
comparison between theory and experiment.

Let us first illustrate in some detail the procedure normally adopted to extract
$(\Gamma_n/\Gamma_p)^{\rm exp}$ from measurements of $N_{nn}/N_{np}$
\cite{ga03,ga04}. Each one of the non--mesonic weak decay channel takes
place by the emission of two or three primary nucleons.
These nucleons propagate within the nuclear environment and cannot be measured.
The strong interactions with the
surrounding nucleons can change the charge and the
energy--momentum of the primary nucleons; some of them can be absorbed
by the medium and the emission of additional (secondary) nucleons can
occur as well. All these processes are generically designated as
final state interactions (FSI): they do not have to be included when
calculating the decay rates, but the observable nucleon spectra,
i.e., $N_{nn}$ and $N_{np}$, are crucially affected by them.
One has to emphasize that,
on the contrary, baryon--baryon short range correlations in both the
initial and the final states as well as mean field
effects on the single particle wave--functions are genuine
contributions to the decay rates.

FSI pertain to the same quantum--mechanical problem which starts with the $\Lambda$
decay and ends with the detection of the particles emitted by the hypernucleus.
In a strict quantum--mechanical scheme,
FSI cannot thus be disentangled from the weak interaction part of the problem:
this is an analogous way of expressing the fact that the weak decay rates
are not measurable.
However, up to now FSI have been simulated by means of semi--classical
models, i.e., by intranuclear cascade codes (INC)~\cite{ra97}
acting after the weak decay, thus losing quantum--mechanical coherence.
In such INC analyses, both one-- and two--nucleon induced decays are included
as inputs and one proceeds to fit $N_{nn}/N_{np}$ data in order to determine
the value of $(\Gamma_n/\Gamma_p)^{\rm exp}$. Technically, this is
achieved by applying Eq.~(16) of Ref.~\cite{ga04}
(see also Eqs.~(1) and (2) of Ref.~\cite{ga03} and Eq.~(3.14)
of Ref.~\cite{Ba06}), which is an exact relation only neglecting
quantum coherence among the final, observable nucleons.
Note that such a procedure also requires a theoretical estimate for the
ratio $\Gamma_2/\Gamma_1$. In other words, present nucleon--nucleon
coincidence data only allows us to determine a correlation property between
$\Gamma_n/\Gamma_p$ and $\Gamma_2/\Gamma_1$.

In general terms, one could wonder if it is possible to identify
those quantum--mechanical contributions whose classical
limit leads to a factorization between the weak decay process and
the INC rescattering.
This is a relevant question since in the theoretical evaluation of the
non--mesonic decay rates FSI contributions must not be included;
one indeed aims to extract the contribution of the elementary
$\Lambda N\to nN$ and $\Lambda nn\to nNN$ processes
by studying hypernuclear decay.
Unfortunately, the above question does not seem to have a simple solution.
Although we make here some considerations about this point,
we believe that a complete answer to it goes beyond the present contribution.

\begin{figure}[t]
\centerline{\includegraphics[scale=0.63]{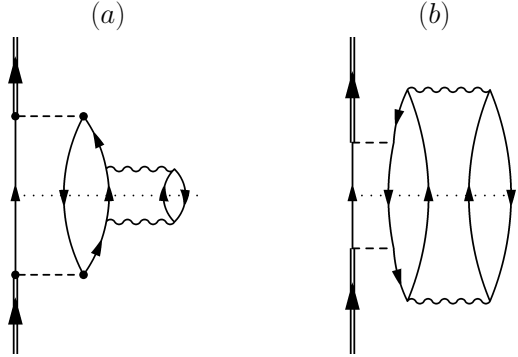}}
\caption{Goldstone diagrams for FSI $(a)$ and $2p2h$ GSC
contributions $(b)$ for three nucleon emission.
The dashed and wavy lines stand for the potentials $V^{\Lambda N \to NN}$
and $V^{NN}$, respectively.
The diagram $(a)$ has poles on the $2p1h$ and $3p2h$ configurations, while $(b)$
has a single pole on the $3p2h$ configuration. For the present discussion
we only consider the $3p2h$ poles indicated by the dotted lines.}
\label{fsgs2p2h}
\end{figure}

Let us illustrate, by using an example, the nature of the problem.
Consider the $\Lambda$ self--energy
diagram $(a)$ of Fig.~\ref{fsgs2p2h}. This is
a (time--ordered) Goldstone diagram where the weak transition potential
$V^{\Lambda N \to NN}$, which is a two--body operator,
produces an intermediate $2p1h$ configuration;
afterwards, the action of the nucleon--nucleon strong interaction
$V^{NN}$ creates a further $1p1h$ pair and leads to a $3p2h$ final state.
In terms of amplitudes, $V^{\Lambda N \to NN}$ produces two
nucleons, one of which then strongly interacts with another
nucleon, ending in the emission of three nucleons.
Since the potential $V^{NN}$ acts after $V^{\Lambda N \to NN}$,
diagram $(a)$ contains a FSI effect and we argue that
it must not be included when evaluating the non--mesonic decay rate.
Note that the idea of an interaction taking place after or before another one is
a valid statement here as we are working with Goldstone diagrams.
On the contrary, diagram $(b)$ of Fig.~\ref{fsgs2p2h}
represents a ground state correlation (GSC) effect. It corresponds
to an amplitude in which the $\Lambda$ decays by interacting with a
correlated nucleon pair. Since the nucleon--nucleon interaction
takes place before the action of the weak transition,
this diagram must be considered when evaluating
the decay rate $\Gamma_{2}$.

Note also that the Goldstone diagrams $(a)$ and $(b)$ are two
different time orderings of the same Feynman diagram.
If $\Gamma_{2}$ were an observable, it would have to be evaluated by means of
Feynman rather than Goldstone diagrams; both diagrams $(a)$ and $(b)$
would contribute to $\Gamma_2$. These diagrams
must actually be taken into account when evaluating the observable nucleon spectra.
However, here we argue that, since $\Gamma_2$ is not an observable,
some of the Goldstone diagram should not be
included in the theoretical definition of this rate.
The class of diagrams that does not contribute to $\Gamma_2$
depends on the definition one adopts for FSI. Our definition
leaves aside those Goldstone diagrams, like diagram $(a)$ in Fig.~\ref{fsgs2p2h},
in which at least one nucleon--nucleon
interaction takes place after the weak transition potential.
If on the other hand one were to include diagram $(b)$ in the
calculation of the widths, then it would not be clear how to
identify the diagrams incorporating FSI effects.

A similar analysis to the previous one holds for the one--nucleon induced rates.
Summarizing, we assume that one-- and two--nucleon induced decay widths, which are not
observables, are interpreted in terms of Goldstone diagrams in which no FSI effect is
present. All the Goldstone diagrams in which at least one nucleon--nucleon interaction
takes place after the weak transition potential must not be included when evaluating the
decay rates. Any Goldstone diagram representing a GSC is instead a genuine contribution
to the rates. In the calculation of the observable nucleon spectra, a description in
terms of Feynman diagrams must instead be employed.

\section{Many--body terms in the non--mesonic decay rates}
\label{miss}
Let us consider the one and two--body induced non--mesonic weak decay
width for a $\Lambda$--hyperon with four--momentum $k=(k_0,\v{k})$
inside infinite nuclear matter with Fermi momentum
$k_F$. In a schematic way, one can write:
\begin{equation}
\label{decw}
\Gamma_{1 \, (2)}(k,k_{F}) = \sum_{f} \,
 |\bra{f} V^{\Lambda N\to NN} \ket{0}_{k_{F}}|^{2}
\delta (E_{f}-E_{0})~,
\end{equation}
where $\ket{0}_{k_{F}}$ and $\ket{f}$ are the initial
hypernuclear ground state (whose energy is $E_0$)
and the possible $2p1h$ or $3p2h$ final states, respectively.
The $2p1h$ ($3p2h$) final states define $\Gamma_{1}$ ($\Gamma_{2}$).
The final state energy is $E_f$ and $V^{\Lambda N\to NN}$ is the
two--body weak transition potential.

The decay rates for a finite hypernucleus are obtained
by the local density approximation~\cite{os85},
i.e., after averaging the above partial width over the $\Lambda$
momentum distribution in the considered hypernucleus,
$|\widetilde{\psi}_{\Lambda}(\v{k})|^2$, and over the local Fermi momentum,
$k_{F}(r) = \{3 \pi^{2} \rho(r)/2\}^{1/3}$,
$\rho(r)$ being the density profile of the hypernuclear core.
One thus has:
\begin{equation}
\label{decwpar3}
\Gamma_{1 \, (2)} = \int d \v{k} \, |\widetilde{\psi}_{\Lambda}(\v{k})|^2
\int d \v{r} \, |\psi_{\Lambda}(\v{r})|^2
\Gamma_{1 \, (2)}(\v{k},k_{F}(r))~,
\end{equation}
where for $\psi_{\Lambda}(\v{r})$, the Fourier transform of
$\widetilde{\psi}_{\Lambda}(\v{k})$, we
adopt the $1s_{1/2}$ harmonic oscillator wave--function with
frequency $\hbar \omega$ ($=10.8$ MeV for $^{12}_\Lambda$C)
adjusted to the experimental energy
separation between the $s$ and $p$ $\Lambda$--levels in the
considered hypernucleus.
The $\Lambda$ total energy in Eqs.~(\ref{decw}) and (\ref{decwpar3})
is given by $k_{0}=m_\Lambda+\v{k}^2/(2 m_\Lambda)+V_{\Lambda}$,
$V_\Lambda$ ($=-10.8$ MeV for $^{12}_\Lambda$C)
being a binding energy term.

\begin{figure}[t]
\centerline{\includegraphics[scale=0.63]{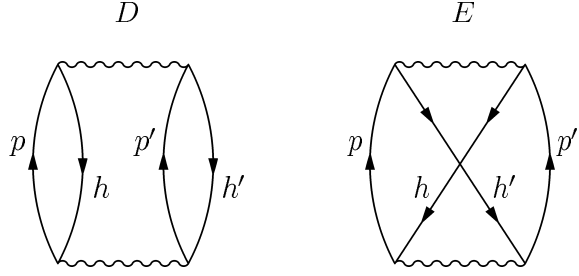}}
\caption{Direct (D) and exchange (E) Goldstone diagrams for the $2p2h$ GSC
induced by the nuclear residual interaction $V^{NN}$.}
\label{gs2p2h}
\end{figure}

Since $V^{\Lambda N\to NN}$ is a two--body operator,
the emission of two nucleons is originated either from the Hartree--Fock
vacuum or from GSC induced by the nucleon--nucleon interaction.
At variance, the emission of three nucleons can be
only achieved when $V^{\Lambda N\to NN}$ acts over a GSC.
It is therefore convenient to introduce the following hypernuclear ground
state wave--function \cite{ba09}:
\begin{equation}
\label{gstate}
\ket{0}_{k_{F}}=\mathcal{N}(k_{F}) \, \left(\ket{\;}
- \sum_{p, h, p', h'} \,
\frac{\bra{p h p' h'} V^{N N} \ket{\;}_{D+E}}
{\varepsilon_{p}-\varepsilon_{h}+\varepsilon_{p'}-\varepsilon_{h'}} \,
\ket{p h p' h'}\right) \otimes \ket{p_{\Lambda}}~,
\end{equation}
where $\ket{\;}$ is the uncorrelated core ground state wave--function,
i.e., the Hartree--Fock vacuum, while the second term in the rhs
represents $2p2h$ correlations and contains both direct ($D$) and exchange
($E$) matrix elements of the nuclear residual interaction $V^{N N}$.
Besides, $\ket{p_{\Lambda}}$ is the normalized state of the $\Lambda$,
the particle and hole energies are denoted by $\varepsilon_{i}$ and:
\begin{equation}
\label{norconst}
\mathcal{N}(k_{F})=\left( 1 + \sum_{p,
h, p', h'} \, \left|\frac{\bra{p h p' h'}
V^{N N} \ket{\;}_{D+E}}
{\varepsilon_{p}-\varepsilon_{h}+\varepsilon_{p'}-\varepsilon_{h'}}
\right|^{2} \, \right)^{-1/2}
\end{equation}
is the ground state normalization function.
The particular labeling of Eqs.~(\ref{gstate}) and (\ref{norconst})
is explained in Fig.~\ref{gs2p2h}. The explicit expression for
$\mathcal{N}(k_{F})$ is given in Ref.~\cite{ba09b}.

By inserting Eq.~(\ref{gstate}) into Eq.~(\ref{decw}),
for $\Gamma_{1}$ one obtains:
\begin{eqnarray}
\label{decw1}
\Gamma_{1}(\v{k},k_{F})  & = & \mathcal{N}^{\, 2}(k_{F})
\sum_{f} \, \delta (E_{f}-E_{0}) \;
\left|\bra{f} V^{\Lambda N\to NN} \ket{p_{\Lambda}}_{D+E}
\phantom{\frac{A^A}{B^A}} \right. \\
&& \left. -\sum_{p, h, p', h'} \,
\bra{f} V^{\Lambda N\to NN} \ket{p h p' h'; \, p_{\Lambda}}_{D+E}
\frac{\bra{p h p' h'; \, p_{\Lambda}} V^{N N} \ket{p_{\Lambda}}_{D+E}}
{\varepsilon_{p}-\varepsilon_{h}+\varepsilon_{p'}-\varepsilon_{h'}}\right|^2~, \nonumber
\end{eqnarray}
the final states $\ket{f}$ being restricted to $2p1h$ states.
For $\Gamma_{2}$ one has:
\begin{eqnarray}
\label{decw2}
\Gamma_{2}(\v{k},k_{F})  & = & \mathcal{N}^{\, 2}(k_{F})
\sum_{f} \, \delta (E_{f}-E_{0}) \\
&&\times \left| \sum_{p, h, p', h'} \,
\bra{f} V^{\Lambda N\to NN} \ket{p h p' h'; \, p_{\Lambda}}_{D+E}
\frac{\bra{p h p' h'; \, p_{\Lambda}} V^{N N} \ket{p_{\Lambda}}_{D+E}}
{\varepsilon_{p}-\varepsilon_{h}+\varepsilon_{p'}-\varepsilon_{h'}}\right|^{2}~, \nonumber
\end{eqnarray}
where the final states are given by $3p2h$ states.
Note that all the matrix elements of $V^{NN}$ and
$V^{\Lambda N\to NN}$ appear in the antisymmetrized form.

Let us focus now on the kind of diagrams contributing to $\Gamma_{1}$ and $\Gamma_{2}$.
This discussion is done in terms of transition amplitudes rather than
self--energies.
\begin{figure}[t]
\centerline{\includegraphics[scale=0.63]{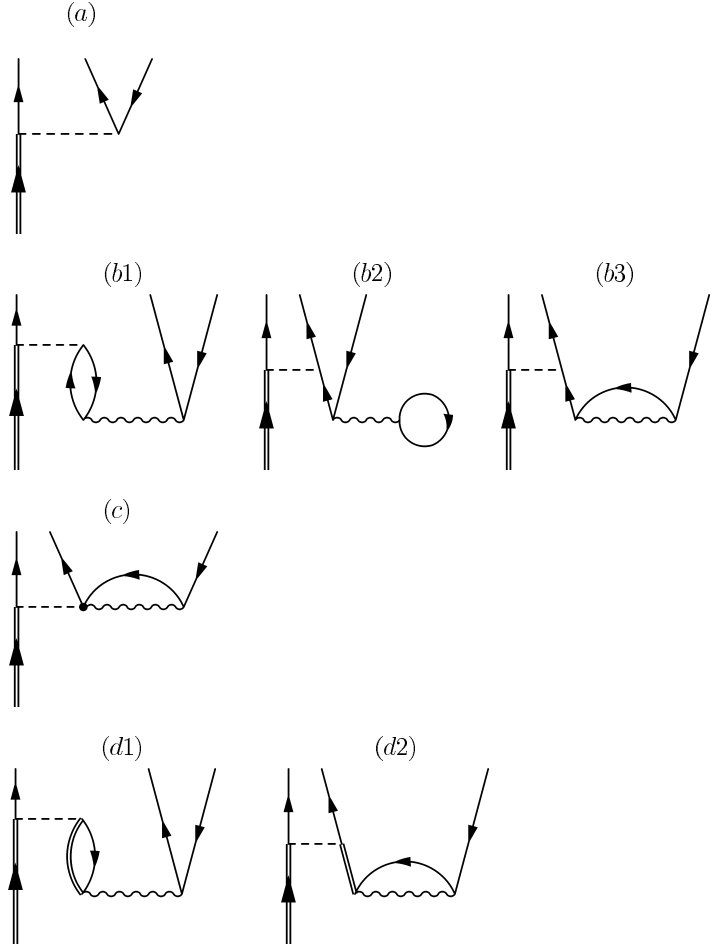}}
\caption{Transition amplitudes contributing to $\Gamma_{1}$.
A double--line (without arrow) represents the $\Delta(1232)$ resonance.}
\label{gam1gsc}
\end{figure}
In Fig.~\ref{gam1gsc} we report some of the most
representative transition amplitudes which contribute to
$\Gamma_{1}$. All diagrams but $(a)$ are originated by a GSC.
Only the contribution of diagram $(a)$ to
$\Gamma_{1}$ has been calculated microscopically up to now.
The line $(b)$ represents typical $2p2h$ correlations. The contribution
$(c)$ is a contact term involving a $\pi \pi NN$ strong vertex,
while line $(d)$ represents the contribution of the $\Delta(1232)$ resonance.
It should be mentioned that there has been a great deal of controversy
around the theoretical determination of the $\Gamma_{n}/\Gamma_{p}$ ratio
and the challenging comparison with data.
In these discussions, all theoretical efforts have been devoted to the $(a)$
term only; the remaining ones have simply been ignored.

\begin{figure}[t]
\centerline{\includegraphics[scale=0.63]{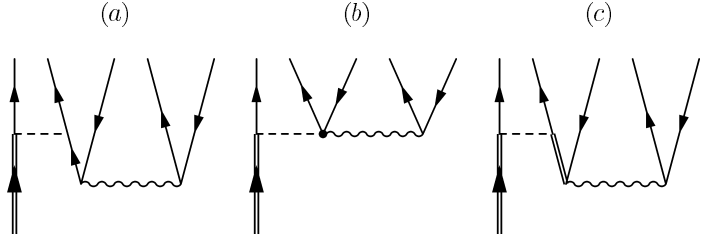}}
\caption{Transition amplitudes contributing to $\Gamma_{2}$.}
\label{gam2gsc}
\end{figure}

\begin{figure}[t]
\centerline{\includegraphics[scale=0.42]{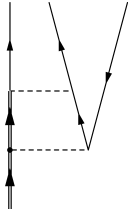}}
\caption{Transition amplitude contributing to $\Gamma_{1}$
and involving a strong interaction $V^{\Lambda N}$ between the hyperon and a
$1p1h$ pair.}
\label{gam1-cor}
\end{figure}

A similar analysis can be done for $\Gamma_{2}$ starting from
the amplitudes of Fig.~\ref{gam2gsc}.
Again, only the $(a)$ term has been evaluated up to now
in microscopic calculations \cite{ba04}.
The graphs in Figs.~\ref{gam1gsc} and \ref{gam2gsc} are only representative cases.
For instance, also the amplitude of Fig.~\ref{gam1-cor}
should be included when calculating $\Gamma_1$. Unlike the other amplitudes
of Figures.~\ref{gam1gsc} and \ref{gam2gsc},
the one in Fig.~\ref{gam1-cor} involves a strong interaction $V^{\Lambda N}$
between the $\Lambda$ and a $1p1h$ pair (i.e., a $1p1h$ GSC) and then
the usual action of the weak transition potential. Apart from the explicit calculation,
such a contribution could in
principle be included in an effective way through the calculation of diagram $(a)$
of Fig.~\ref{gam1gsc} with a suitably chosen weak transition potential
$V^{\Lambda N\to NN}$.
However, based on the absence of isovector--meson exchange in the strong potential
$V^{\Lambda N}$, one may anticipate a small effect of this amplitude.
Other amplitudes will provide important contributions.
In the graph $(a)$ of Fig.~\ref{gam2gsc} the weak transition potential
can also be connected to a hole line \cite{ba04}. In addition,
since $V^{N N}$ and $V^{\Lambda N\to NN}$ are two--body operators
whose matrix elements are antisymmetrized,
Pauli exchange terms must be considered as well \cite{ba09b}.

All the graphs in Figs.~\ref{gam1gsc} and \ref{gam2gsc} have the same
initial state, which is the hypernuclear ground state.
The final state of the graphs in Fig.~\ref{gam1gsc} (Fig.~\ref{gam2gsc}) is
a $2p1h$ ($3p2h$) state. To obtain the various
decay width, all graphs representing transitions amplitudes
with the same initial and final states are added and then
squared. For instance, from Fig.~\ref{gam2gsc} one obtains
a total of six direct diagrams: the square of each individual
amplitude plus the three interference terms. For
the amplitudes in Fig.~\ref{gam1gsc}
there is a total of twenty--eight different direct terms. In addition,
antisymmetrization considerably increases the amount of diagrams.
From our previous works it is clear to us
that a full microscopic evaluation of each term is
mandatory for several reasons. First, a raw estimation
of a remarkable amount of different diagrams makes the final result
quite uncertain. Secondly, there is no ground to
evaluate differently the diagrams originated from
Fig.~\ref{gam1gsc} and those from Fig.~\ref{gam2gsc}: once a microscopic
calculation is performed for the square of diagrams $(a)$
of Figs.~\ref{gam1gsc} and \ref{gam2gsc},
the same should be done for the remaining contributions,
which are all leading order GSC contributions.

In the present work, as a further step towards the calculation of the
whole set of diagrams relevant for the non--mesonic decay, the one--nucleon
induced widths originated from the sum of the transition amplitudes $(a)$ plus $(b1)$ of
Fig.~\ref{gam1gsc} are evaluated for the first time. Accordingly,
the two--nucleon induced rates are instead obtained from the amplitude
$(a)$ of Fig.~\ref{gam2gsc} by following Ref.~\cite{ba09b}.
Antisymmetrization is coherently applied to all contributions.
Before proceeding with the formal derivation of the decay widths,
in the next Section
we first point out additional observations on the evaluation of
the decay rates and on previous, related work.

\section{Further considerations on the evaluation of the weak decay rates}
\label{fsi}
Let us start this discussion by paying attention to the twofold effect of the
nuclear residual interaction $V^{NN}$ within the matrix elements of Eq.~(\ref{decw}).
When $V^{NN}$ acts on the uncorrelated hypernuclear ground state $\ket{\,}$,
as in Eq.~(\ref{gstate}), one has a GSC.
Alternatively, $V^{NN}$ may introduce medium effects on the
weak transition potential $V^{\Lambda N \to NN}$. Both
effects must be taken into account when calculating the decay rates.
In addition, $V^{NN}$ may modify the final states $\ket{f}=\ket{2p1h}$
or $\ket{3p2h}$: for instance, acting on a $\ket{2p1h}$ final state,
it can produce a $\ket{3p2h}$ state, as in Fig.~\ref{fsgs2p2h}(a); this results
in a FSI which does not contribute to Eq.~(\ref{decw}).

Concerning the medium effects previously mentioned, let us
discuss some aspects of the work of Ref.~\cite{Ji01}. Here, $V^{NN}$
introduces medium modifications on the mesons propagators appearing
in $V^{\Lambda N \to NN}$ through the direct part of the RPA (ring approximation):
schematically, in our scheme
one simply has to replace $V^{\Lambda N \to NN}$ with
$\widetilde{V}^{\Lambda N \to NN}=V^{\Lambda N \to NN}/|1-\Pi V^{NN}|$,
where the polarization propagator $\Pi$ contains $1p1h$ and $1\Delta 1h$
contributions in Ref.~\cite{Ji01}.
Note that, since only the absolute value of the ring propagator is kept,
the modified weak transition potential remains a real function.
This approach thus represents a refinement of the weak transition potential
and is consistent with Eq.~(\ref{decw}).

\begin{figure}[t]
\centerline{\includegraphics[scale=0.50]{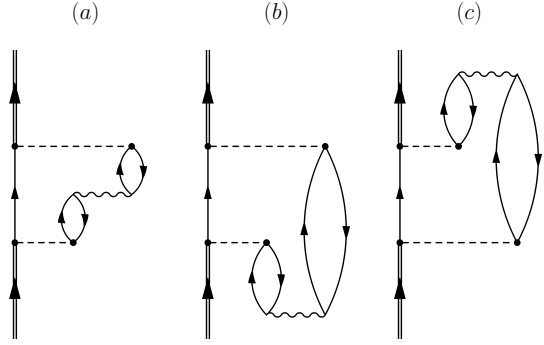}}
\caption{Goldstone diagrams for FSI $(a)$ and $2p2h$ GSC
contributions $(b)$ and $(c)$ leading to two nucleon emission.
}
\label{figfsiab}
\end{figure}

We emphasize that the mere use of diagrams when discussing
the formalism developed in Ref.~\cite{Ji01} or the present one
could be misleading. For the approximation considered in Ref.~\cite{Ji01},
in Eq.~(\ref{decw}) one has to employ
the matrix element $\bra{f} \widetilde{V}^{\Lambda N \to NN} \ket{0}_{D+E}$
of the modified weak transition potential (the corresponding direct and
exchange self--energy diagrams are shown in Fig.~2 in
Ref.~\cite{Ji01}). By making an expansion of the square of this matrix element
in the ring series, the two terms at first order in $V^{NN}$ correspond to
a self--energy contribution
which matches exactly with the $(a)$ diagram in Fig.~\ref{figfsiab},
where the final state $\ket{f}=\ket{2p1h}$ corresponds either to the upper or
the lower bubble.
Nevertheless, the same diagram could also be associated to the direct part
of the following product of matrix elements:
$\bra{f} V^{NN} \ket{i}_{D+E} \, \bra{i} V^{\Lambda N \to NN} \ket{0}_{D+E}$,
where $\ket{i}$ is a $2p1h$ intermediate configuration.
But, since this product contains a FSI, it
is not a correct contribution to the decay rates of Eq.~(\ref{decw}).
Antisymmetry of this product of matrix elements
gives rise to a total of eight self--energy
diagrams, which are shown in Fig.~4 in Ref.~\cite{ba07b}
and used there to calculate the (observable) spectra of the
non--mesonic weak decay nucleons.
From the analytical point of view, the product
$\bra{f} V^{NN} \ket{i}_{D+E} \, \bra{i} V^{\Lambda N \to NN} \ket{0}_{D+E}$
is clearly different from the term at first
order in $V^{NN}$ entering
$\bra{f} \widetilde{V}^{\Lambda N \to NN} \ket{0}_{D+E}$.
When the comparison is done using the full set of direct plus exchange diagrams,
FSI and the medium modifications on
the weak transition potential are manifestly different effects.
Only the latter can be included in the calculation of the decay rates.

As a final remark for this section, we observe that the amplitudes
$(a)$ and $(b1)$ of Fig.~\ref{gam1gsc} produce the self--energy
diagrams $(b)$ and $(c)$ of Fig.~\ref{figfsiab}.
They are GSC terms and thus contribute to the decay rates. Conversely,
the $(a)$ diagram of Fig.~\ref{figfsiab} must be left aside in the
calculation, unless one
considers it as a medium modification on the weak transition potential
(but then, other medium modification contributions should be considered
simultaneously), as done in Ref.~\cite{Ji01}.
The Goldstone diagrams of Fig.~\ref{figfsiab} are the three possible
time orderings of the same Feynman diagram. Again, we stress that
the fact that one out of
three diagrams in Fig.~\ref{figfsiab} will not be included in our
calculation of the decay rates makes sense since these rates
are not observables and thus do not have to be described
by Feynman diagrams.

\section{Formal derivation of the decay rates
$\bf{\Gamma_{n}}$ and $\bf{\Gamma_{p}}$ including GSC}
\label{gngpgsc}
In Fig.~\ref{gam1gsc} we have shown a set of amplitudes
which contribute to the decay rate $\Gamma_{1}$ of Eq.~(\ref{decw}).
Only the amplitude $(a)$ has been
evaluated explicitly up to now. In the present work we
extend the microscopic approach to include the amplitude $(b1)$,
which originates from GSC contributions that we expect to be important.

Before proceeding with the derivation of decay widths, it is
convenient to give the expressions for the potentials.
The weak transition potential $V^{\Lambda N \to NN}$ and the
nuclear residual interaction $V^{NN}$ read:
\begin{equation}
\label{intlnnn}
V^{\Lambda N\to NN (NN)} (q) = \sum_{\tau_{\Lambda (N)}=0,1}
 {\cal O}_{\tau_{\Lambda (N)}}
{\cal V}_{\tau_{\Lambda (N)}}^{\Lambda N \to NN (NN)} (q)~,
\end{equation}
where the isospin dependence is given by
\begin{eqnarray}
\label{isos} {\cal O}_{\tau_{\Lambda (N)}} =~~~~~
\left\{
\begin{array}{c}1~~~~~\mbox{for}~~\tau_{\Lambda (N)}=0\\
  \v{\tau}_1 \cdot \v{\tau}_2~~\mbox{for}~~\tau_{\Lambda (N)}=1~.
\end{array}\right.
\end{eqnarray}
The values $0$ and $1$ for $\tau_{\Lambda (N)}$ refer to the isoscalar
and isovector parts of the interactions, respectively.
The spin and momentum dependence of the weak transition potential is
given by:
\begin{eqnarray}
\label{intln}
{\cal V}_{\tau_{\Lambda}}^{\Lambda N \to NN} (q) &
= &  (G_F m_{\pi}^2)  \; \{
S_{\tau_{\Lambda}}(q)  \; \v{\sigma}_1 \cdot \v{\hat{q}} +
S'_{\tau_{\Lambda}}(q)  \; \v{\sigma}_2 \cdot \v{\hat{q}} +
P_{C, \tau_{\Lambda}}(q) \\
& &
+ P_{L, \tau_{\Lambda}}(q)
\v{\sigma}_1 \cdot \v{\hat{q}} \; \v{\sigma}_2 \cdot
\v{\hat{q}}  +
P_{T, \tau_{\Lambda}}(q)
(\v{\sigma}_1 \times \v{\hat{q}}) \cdot (\v{\sigma}_2 \times
\v{\hat{q}}) \nonumber \\
& &
+i S_{V, \tau_{\Lambda}}(q)
\v{(\sigma}_1 \times \v{\sigma}_2) \cdot
\v{\hat{q}} \}~, \nonumber
\end{eqnarray}
where the functions $S_{\tau_{\Lambda}}(q)$, $S'_{\tau_{\Lambda}}(q)$,
$P_{C, \tau_{\Lambda}}(q)$, $P_{L,\tau_{\Lambda}}(q)$,
$P_{T, \tau_{\Lambda}}(q)$ and $S_{V, \tau_{\Lambda}}(q)$,
which include short range correlations,
are adjusted to reproduce any weak transition potential.

The corresponding expression for the nuclear residual interaction is given by:
\begin{eqnarray}
\label{intnn}
{\cal V}_{\tau_N}^{N N} (q)
& = & \frac{f_{\pi}^2}{m_{\pi}^2}  \; \{
{\cal V}_{C, \,\tau_{N}}(q) +
{\cal V}_{L, \, \tau_{N}}(q)
\v{\sigma}_1 \cdot \v{\hat{q}} \; \v{\sigma}_2 \cdot
\v{\hat{q}} \\
& & + {\cal V}_{T, \, \tau_{N}}(q)
(\v{\sigma}_1 \times \v{\hat{q}}) \cdot (\v{\sigma}_2 \times
\v{\hat{q}}) \}~, \nonumber
\end{eqnarray}
where the functions ${\cal V}_{C, \,\tau_{N}}(q)$,
${\cal V}_{L, \, \tau_{N}}(q)$ and
${\cal V}_{T, \, \tau_{N}}(q)$ are also adjusted to reproduce
any nuclear residual interaction.

In particular, $V^{\Lambda N \to NN}$ is represented by the exchange of the $\pi$,
$\eta$, $K$, $\rho$, $\omega$ and $K^*$ mesons, within the formulation
of Ref.~\cite{pa97}, with strong
coupling constants and cut--off parameters deduced from
the Nijmegen soft--core interaction NSC97f of
Ref.~\cite{st99}. For $V^{N N}$ we have used a Bonn
potential~\cite{ma87} in the framework of the parametrization
presented in Ref.~\cite{br96}, which contains the exchange of $\pi$,
$\rho$, $\sigma$ and $\omega$ mesons.

We give now explicit expressions for the partial decay
width $\Gamma_{1}(\v{k},k_{F})$ of Eq.~(\ref{decw1}), which
for convenience is expressed in terms of its isospin components
$\Gamma_{n}(\v{k},k_{F})$ and $\Gamma_{p}(\v{k},k_{F})$.
Let us first rewrite
Eq.~(\ref{decw1}) as follows:
\begin{equation}
\label{0gsc}
\Gamma_{n \, (p)}(\v{k},k_{F})=
\Gamma^{0}_{n \, (p)}(\v{k},k_{F})+\Gamma^{0-\rm GSC}_{n \, (p)}(\v{k},k_{F})
+\Gamma^{\rm GSC}_{n \, (p)}(\v{k},k_{F})~,
\end{equation}
where:
\begin{eqnarray}
\label{decw10}
\Gamma^{0}_{n \, (p)}(\v{k},k_{F})  & = & \mathcal{N}^{\, 2}(k_{F})
\sum_{f} \, \delta (E_{f}-E_{0}) \;
\left|\bra{f} V^{\Lambda N\to NN} \ket{p_{\Lambda}}_{D+E}\right|^{2}~, \\
\label{decw10gsc}
\Gamma^{0-\rm GSC}_{n \, (p)}(\v{k},k_{F})  & = & - 2 \mathcal{N}^{\, 2}(k_{F})
\sum_{f} \sum_{p, h, p', h'} \, \delta (E_{f}-E_{0}) \;
\bra{p_{\Lambda}} (V^{\Lambda N\to NN})^{\dagger} \ket{f}_{D+E} \nonumber \\
&&
\times \bra{f} V^{\Lambda N\to NN} \ket{p h p' h'; \, p_{\Lambda}}_{D+E}
\frac{\bra{p h p' h'; \, p_{\Lambda}} V^{N N} \ket{p_{\Lambda}}_{D+E}}
{\varepsilon_{p}-\varepsilon_{h}+\varepsilon_{p'}-\varepsilon_{h'}}~,\nonumber\\
\label{decw1gsc}
\Gamma^{\rm GSC}_{n \, (p)}(\v{k},k_{F})  & = & \mathcal{N}^{\, 2}(k_{F})
\sum_{f} \sum_{p, h, p', h'} \, \delta (E_{f}-E_{0}) \;
\left|\bra{f} V^{\Lambda N\to NN} \ket{p h p' h'; \, p_{\Lambda}}_{D+E}
\phantom{\frac{A^A}{A^A}} \right. \nonumber \\
&& \left. \times \frac{\bra{p h p' h'; \, p_{\Lambda}} V^{N N} \ket{p_{\Lambda}}_{D+E}}
{\varepsilon_{p}-\varepsilon_{h}+\varepsilon_{p'}-\varepsilon_{h'}}\right|^{2}~. \nonumber
\end{eqnarray}
The first component, $\Gamma^{0}_{n \, (p)}$, is the contribution from the
uncorrelated hypernuclear ground state,
the third one, $\Gamma^{\rm GSC}_{n \, (p)}$, result from
ground state correlations,
while $\Gamma^{0-\rm GSC}_{n \, (p)}$ is the interference term between
correlated and uncorrelated ground states.

It is now convenient to consider the following decomposition,
dictated by the isospin quantum number:
\begin{eqnarray}
\label{rpa2}
\Gamma^{0}_{n \, (p)}(\v{k},k_{F})& =
&\sum_{P,Q=D, \, E} \, \Gamma^{PQ}_{n\, (p)}(\v{k},k_{F}) \\
&=&\sum_{P,Q=D, \, E} \, \sum_{\tau_{\Lambda'}, \tau_{\Lambda}=0,1}
{\cal T}^{PQ}_{\tau_{\Lambda'} \tau_{\Lambda}, \; n \, (p)} \;
\Gamma^{PQ}_{\tau_{\Lambda'} \tau_{\Lambda}}(\v{k},k_{F})~, \nonumber \\
\Gamma^{0-\rm GSC}_{n \, (p)}(\v{k},k_{F})& =
&\sum_{P,Q,Q'=D, \, E} \, \Gamma^{PQQ'}_{n\, (p)}(\v{k},k_{F}) \nonumber \\
&=&\sum_{P,Q,Q'=D, \, E} \,\sum_{\tau_{\Lambda'}, \tau_{\Lambda}, \tau_{N}=0,1}
{\cal T}^{PQQ'}_{\tau_{\Lambda'} \tau_{\Lambda} \tau_{N}, \; n \, (p)} \;
\Gamma^{PQQ'}_{\tau_{\Lambda'} \tau_{\Lambda} \tau_{N}}(\v{k},k_{F})~, \nonumber \\
\Gamma^{\rm GSC}_{n \, (p)}(\v{k},k_{F})& =
&\sum_{P',P,Q,Q'=D, \, E} \, \Gamma^{P'PQQ'}_{n\, (p)}(\v{k},k_{F}) \nonumber \\
&=&\sum_{P',P,Q,Q'=D, \, E} \,\sum_{\tau_{N'}, \tau_{\Lambda'}, \tau_{\Lambda}, \tau_{N}=0,1}
{\cal T}^{P'PQQ'}_{\tau_{N'} \tau_{\Lambda'} \tau_{\Lambda} \tau_{N}, \; n \, (p)} \;
\Gamma^{P'PQQ'}_{\tau_{N'} \tau_{\Lambda'} \tau_{\Lambda} \tau_{N}}(\v{k},k_{F})~, \nonumber
\end{eqnarray}
where $P', P, \, Q, \, Q'=$ $D$ or $E$
refer to the direct or exchange character of the matrix elements
of Eq.~(\ref{decw10}). The isospin factors are given by:
\begin{eqnarray}
\label{rpa3}
{\cal T}^{PQ}_{\tau_{\Lambda'} \tau_{\Lambda}, \; n \, (p)} & = &
\sum_{f,\, \rm isospin} \, \bra{t_{\Lambda}} {\cal O}_{\tau_{\Lambda'}} \ket{f}_{P}
\bra{f} {\cal O}_{\tau_{\Lambda}} \ket{t_{\Lambda}}_{Q}~, \nonumber \\
{\cal T}^{PQQ'}_{\tau_{\Lambda'} \tau_{\Lambda} \tau_{N}, \; n \, (p)} & = &
\sum_{f,\, \rm isospin} \, \bra{t_{\Lambda}} {\cal O}_{\tau_{\Lambda'}} \ket{f}_{P}
\bra{f} {\cal O}_{\tau_{\Lambda}} \ket{t_{p} t_{h} t_{p'} t_{h'}, t_{\Lambda}}_{Q} \nonumber \\
&&\times \bra{t_{p} t_{h} t_{p'} t_{h'}, t_{\Lambda}}
{\cal O}_{\tau_{N}} \ket{t_{\Lambda}}_{Q'}~, \nonumber \\
{\cal T}^{P'PQQ'}_{\tau_{N'} \tau_{\Lambda'} \tau_{\Lambda} \tau_{N}, \; n \, (p)} & = &
\sum_{f,\, \rm isospin} \, \bra{t_{\Lambda}} {\cal O}_{\tau_{N'}} \ket{t_{\tilde{p}}
t_{\tilde{h}} t_{\tilde{p}'} t_{\tilde{p}'}, t_{\Lambda}}_{P'}
\bra{t_{\tilde{p}} t_{\tilde{h}} t_{\tilde{p}'}
t_{\tilde{p}'}, t_{\Lambda}} {\cal O}_{\tau_{\Lambda'}} \ket{f}_{P} \nonumber \\
&& \times \bra{f} {\cal O}_{\tau_{\Lambda}} \ket{t_{p} t_{h} t_{p'} t_{h'}, t_{\Lambda}}_{Q}
\bra{t_{p} t_{h} t_{p'} t_{h'}, t_{\Lambda}} {\cal O}_{\tau_{N}} \ket{t_{\Lambda}}_{Q'}~, \nonumber
\end{eqnarray}
where the summations run over all the isospin projections $t's$,
with the constrain that the emitted particles are
$nn$ for $\Gamma_{n}$ and $np$ for $\Gamma_{p}$.
For the partial decay widths we instead find:
\begin{eqnarray}
\Gamma^{PQ}_{\tau_{\Lambda'} \tau_{\Lambda}}(\v{k},k_{F})
& = & \mathcal{N}^{\, 2}(k_{F}) \, (-1)^{n} \,
\sum_{f} \, \delta (E_{f}-E_{0}) \\
\label{gampq}
&& \times \bra{p_{\Lambda}} ({\cal V}_{\tau_{\Lambda'}}^{\Lambda N \to NN} (q'))^{\dag} \ket{f}_{P}
\bra{f} {\cal V}_{\tau_{\Lambda}}^{\Lambda N \to NN} (q) \ket{p_{\Lambda}}_{Q}~, \nonumber \\
\label{gampqq}
\Gamma^{PQQ'}_{\tau_{\Lambda'} \tau_{\Lambda} \tau_{N}}(\v{k},k_{F})
& = & - 2 \, \mathcal{N}^{\, 2}(k_{F}) \, (-1)^{n} \,
\sum_{f} \,  \sum_{p, h, p', h'} \, \delta (E_{f}-E_{0}) \\
&& \times \bra{p_{\Lambda}} ({\cal V}_{\tau_{\Lambda'}}^{\Lambda N \to NN} (q'))^{\dag} \ket{f}_{P}
\bra{f} {\cal V}_{\tau_{\Lambda}}^{\Lambda N \to NN} (q)
\ket{p h p' h'; \, p_{\Lambda}}_{Q} \nonumber \\
&&\times \frac{\bra{p h p' h'; \, p_{\Lambda}} {\cal V}_{\tau_N}^{N N} (t) \ket{p_{\Lambda}}_{Q'}}
{\varepsilon_{p}-\varepsilon_{h}+\varepsilon_{p'}-\varepsilon_{h'}}~, \nonumber \\
\label{gamppqq}
\Gamma^{P'PQQ'}_{\tau_{N'} \tau_{\Lambda'} \tau_{\Lambda} \tau_{N} }(\v{k},k_{F})
& = & \mathcal{N}^{\, 2}(k_{F}) \, (-1)^{n} \,
\sum_{f} \,  \sum_{\tilde{p}, \tilde{h}, \tilde{p}', \tilde{h}'} \,
\sum_{p, h, p', h'} \, \delta (E_{f}-E_{0}) \\
&&
\times \frac{\bra{p_{\Lambda}} ({\cal V}_{\tau_{N'}}^{N N}(t'))^{\dag}
\ket{\tilde{p}, \tilde{h}, \tilde{p}', \tilde{h}'; \, p_{\Lambda}}_{P'}}
{\varepsilon_{\tilde{p}}-\varepsilon_{\tilde{h}}+
\varepsilon_{\tilde{p}'}-\varepsilon_{\tilde{h}'}} \nonumber \\
&& \times
\bra{\tilde{p}, \tilde{h}, \tilde{p}', \tilde{h}';
\, p_{\Lambda}} ({\cal V}_{\tau_{\Lambda'}}^{\Lambda N \to NN}(q'))^{\dag}
\ket{f}_{P} \nonumber \\
&& \times \bra{f} {\cal V}_{\tau_{\Lambda}}^{\Lambda N \to NN} (q)
\ket{p h p' h'; \, p_{\Lambda}}_{Q} \nonumber\\
&&\times \frac{\bra{p h p' h'; \, p_{\Lambda}} {\cal V}_{\tau_N}^{N N} (t) \ket{p_{\Lambda}}_{Q'}}
{\varepsilon_{p}-\varepsilon_{h}+\varepsilon_{p'}-\varepsilon_{h'}}~. \nonumber
\end{eqnarray}
Note that the values of the energy--momentum carried by the particles
and holes lines depends on the topology of the corresponding diagram, while
$n$ is the number of crossing between fermionic lines.

\begin{figure}[t]
\centerline{\includegraphics[scale=0.63]{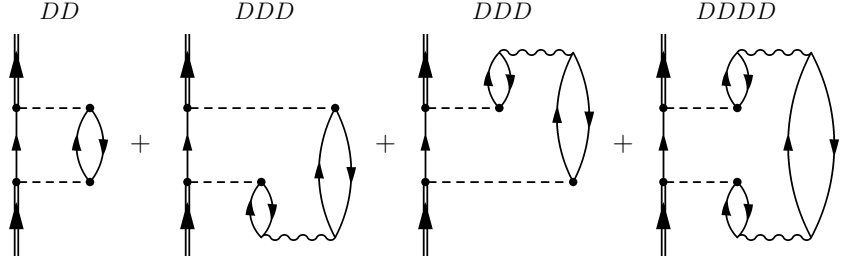}}
\caption{Direct Goldstone diagrams corresponding to the square of the
amplitude sum $(a)+(b1)$ of Fig.~\ref{gam1gsc}. See the decomposition
of Eq.~(\ref{rpa2}).}
\label{dirgsc}
\end{figure}

Let us now apply the above formalism to a model including
the amplitudes $(a)$ and $(b1)$ of Fig.~\ref{gam1gsc}.
Four direct self--energy diagrams correspond to the square
of the amplitude sum $(a)+(b1)$; they are given in Fig.~\ref{dirgsc}.
Note that these diagrams admits a single cut, giving rise to a $2p1h$
final state.
The $DD$ diagram contributes to the partial widths $\Gamma^0_{n\, (p)}$ of
Eq.~(\ref{rpa2}). The two $DDD$ diagrams, which have the same numerical value
and are interferences between the amplitudes $(a)$ and $(b1)$ of Fig.~\ref{gam1gsc},
are included in the partial widths $\Gamma^{0-\rm GSC}_{n\, (p)}$.
Finally, the diagram $DDDD$ contributes to $\Gamma^{\rm GSC}_{n\, (p)}$.
Many exchange diagrams are obtained from the antisymmetrized amplitude sum $(a)+(b1)$:
one $PQ$ exchange diagram is the partner of the $DD$ one of Fig.~\ref{dirgsc};
seven $PQQ'$ exchange diagrams are companions of each one of the $DDD$ ones;
fifteen $P'PQQ'$ exchange diagrams add to the $DDDD$ one.

\begin{figure}[t]
\centerline{\includegraphics[scale=0.63]{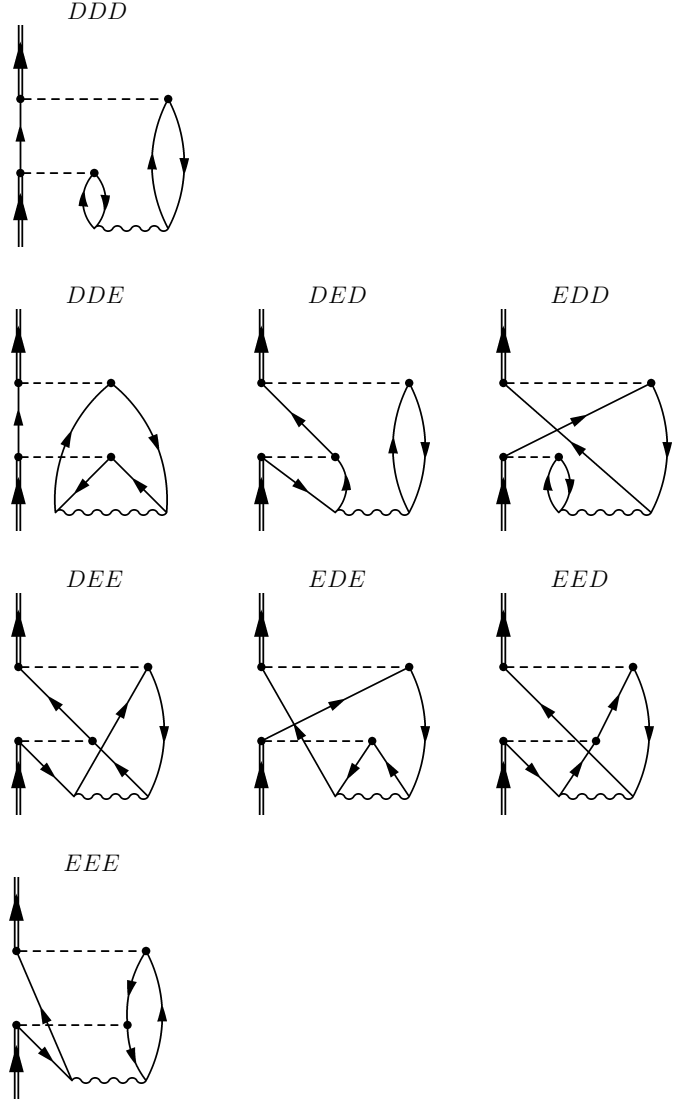}}
\caption{Goldstone diagrams for the partial rates $\Gamma^{PQQ'}_{n \, (p)}$
contributing to Eq.~(\ref{rpa2}).}
\label{antgsc}
\end{figure}

Formal expressions for $\Gamma^{0}_{n \, (p)}$ can be found in Ref.~\cite{ba03}.
The $\Gamma^{PQQ'}_{n \, (p)}$'s contributing to
$\Gamma^{0-\rm GSC}_{n \, (p)}$ (see Eq.~(\ref{rpa2}))
correspond to the diagrams of Fig.~\ref{antgsc}.
By replacing, in Eq.~(\ref{gampqq}),
the sum over momenta by integrals and by performing the
energy integrations and the spin summation, the following expression for
$\Gamma^{PQQ'}_{\tau_{\Lambda'} \tau_{\Lambda} \tau_{N}}$
can be obtained:
\begin{eqnarray}
\label{pqqgsc}
\Gamma^{PQQ'}_{\tau_{\Lambda'} \tau_{\Lambda} \tau_{N}}(\v{k},k_F) & =
& \mathcal{N}^{\, 2}(k_{F})
\frac{1}{4} \frac{(-1)^{n}}{(2 \pi)^8} (G_F m_{\pi}^2)^2
\frac{f_{\pi}^2}{m_{\pi}^2} \\
&&\times \int  \int \int \, d \v{q} \, d \v{h} \, d \v{h}' \;
{\cal W}^{PQQ'}_{\tau_{\Lambda'} \tau_{\Lambda} \tau_{N}}(q, q', t) \nonumber \\
&&\times \Theta(k,q,q',t,h,h',k_{F}) \frac{1}{- \varepsilon^{PQQ'}_{2p2h}} \;
\delta(q_0 - (\varepsilon_{\v{h}'+\v{q}}-\varepsilon_{\v{h}'}))~, \nonumber
\end{eqnarray}
where $q_0=k_0 - \varepsilon_{\v{k}-\v{q}} - V_N$, $V_N$
being the nucleon binding energy, while
the functions ${\cal W}^{PQQ'}_{\tau_{\Lambda'} \tau_{\Lambda} \tau_{N}}(q, q', t)$
and $\Theta(k,q,q',t,h,h',k_{F})$
and the energy denominator $\varepsilon^{PQQ'}_{2p2h}$ are
specific of each $PQQ'$ contribution.
The function ${\cal W}^{PQQ'}_{\tau_{\Lambda'} \tau_{\Lambda} \tau_{N}}(q, q', t)$
contains the momentum dependence of the nuclear residual interaction
and the weak transition potentials and the spin summation, while
$\Theta(k,q,q',t,h,h',k_{F})$ is a product of step functions which defines
the phase space of particles and holes.

In the present section we present the explicit expression for the direct
term $\Gamma^{DDD}_{\tau_{\Lambda'} \tau_{\Lambda} \tau_{N}}$;
the other seven ones are displayed in Appendix~A. We obtain:
\begin{eqnarray}
\label{dddgsc}
\Gamma^{DDD}_{\tau_{\Lambda'} \tau_{\Lambda} \tau_{N}}(\v{k},k_F)  & =
& \mathcal{N}^{\, 2}(k_{F})
\frac{1}{4} \frac{1}{(2 \pi)^8} (G_F m_{\pi}^2)^2
\frac{f_{\pi}^2}{m_{\pi}^2} \\
&&\times \int  \int \int \, d \v{q} \, d \v{h} \, d \v{h}' \;
{\cal W}^{DDD}_{\tau_{\Lambda'} \tau_{\Lambda} \tau_{N}}(q)  \;
\nonumber \\
& &
\times \theta(q_0) \theta(|\v{k}-\v{q}|-k_{F}) \theta(|\v{q}-\v{h}|-k_{F}|)
\theta(k_{F}-|\v{h}|)
\nonumber \\
& & \times \theta(|\v{q}+\v{h}'|-k_{F}|)
\theta(k_{F}-|\v{h}'|) \nonumber \\
& &
\times \frac{1}{-q_{0}-(\varepsilon_{\v{h}-\v{q}}-\varepsilon_{\v{h}})} \;
\delta(q_0 - (\varepsilon_{\v{h}'+\v{q}}-\varepsilon_{\v{h}'}))~. \nonumber
\end{eqnarray}
The expressions for
$\Theta(k,q,q',t,h,h',k_{F})$ and $\varepsilon^{DDD}_{2p2h}$
are self--evident. Moreover:
\begin{eqnarray}
\label{tdir3}
{\cal W}^{DDD}_{\tau_{\Lambda'} \tau_{\Lambda} \tau_{N}}(q) & = &
8 \; \{ [S'_{\tau_{\Lambda'}}(q) S'_{\tau_{\Lambda}}(q)
+ P_{C, \tau_{\Lambda'}}(q) P_{C, \tau_{\Lambda}}(q)] {\cal V}_{C, \,\tau_{N}}(q) \\
&&+ [S_{\tau_{\Lambda'}}(q) S_{\tau_{\Lambda}}(q)+P_{L, \tau_{\Lambda'}}(q)
P_{L, \tau_{\Lambda}}(q)]
 {\cal V}_{L, \, \tau_{N}}(q) \nonumber \\
&&  + 2  \, [S_{V, \tau_{\Lambda'}}(q) S_{V, \tau_{\Lambda}}(q) +
P_{T, \tau_{\Lambda'}}(q) P_{T, \tau_{\Lambda}}(q)]
{\cal V}_{T, \, \tau_{N}}(q) \}~.
\end{eqnarray}
Eq.~(\ref{dddgsc}) can be simplified by introducing the functions:
\[
\label{lind}
{\cal I}(q_{0},\v{q}) =
\frac{-\pi}{(2 \pi)^{3}} \int \, d \v{h}'
\theta(|\v{q}+\v{h}'|-k_{F}|)
\theta(k_{F}-|\v{h}'|)
\delta(q_0 - \varepsilon_{\v{h}'+\v{q}}+\varepsilon_{\v{h}'})~, \nonumber
\]
\begin{equation}
\label{real}
{\cal R}(q_{0},\v{q}) =
\frac{1}{(2 \pi)^{3}} \, {\cal P} \, \int \, d \v{h}
\frac{\theta(|\v{q}-\v{h}|-k_{F}|) \theta(k_{F}-|\v{h}|)}
{q_0 - (\varepsilon_{\v{h}-\v{q}}-\varepsilon_{\v{h}})}~,
\end{equation}
where ${\cal I}(q_{0},\v{q})$ is the imaginary part of the Lindhard function and the
explicit expression for ${\cal R}(q_{0},\v{q})$ is given in Appendix~B.
Therefore:
\begin{eqnarray}
\label{dddgsc2}
\Gamma^{DDD}_{\tau_{\Lambda'} \tau_{\Lambda} \tau_{N}}(\v{k},k_F) & =
& - \, \frac{\mathcal{N}^{\, 2}(k_{F})}{(2 \pi)^3} (G_F m_{\pi}^2)^2
\frac{f_{\pi}^2}{m_{\pi}^2} \, \int  \,
d \v{q} \, \theta(q_0) \theta(|\v{k}-\v{q}|-k_{F}) \\
&& \times \{ [S'_{\tau_{\Lambda'}}(q) S'_{\tau_{\Lambda}}(q) +
P_{C, \tau_{\Lambda'}}(q) P_{C, \tau_{\Lambda}}(q)]
{\cal V}_{C, \,\tau_{N}}(q) \nonumber \\
&&+ [S_{\tau_{\Lambda'}}(q)
S_{\tau_{\Lambda}}(q)+P_{L, \tau_{\Lambda'}}(q) P_{L, \tau_{\Lambda}}(q)]
{\cal V}_{L, \, \tau_{N}}(q) \nonumber \\
&&+ 2  \, [S_{V, \tau_{\Lambda'}}(q) S_{V, \tau_{\Lambda}}(q) +
P_{T, \tau_{\Lambda'}}(q) P_{T, \tau_{\Lambda}}(q)] {\cal V}_{T, \, \tau_{N}}(q)\}
\nonumber \\
&&\times{\cal R}(-q_{0},\v{q}) {\cal I}(q_{0},\v{q})~. \nonumber
\end{eqnarray}

Then one has to perform the isospin summation to obtain
\begin{equation}
\Gamma^{DDD}_{n\, (p)}(\v{k},k_{F})
=\sum_{\tau_{\Lambda'}, \tau_{\Lambda}, \tau_{N}=0,1}
{\cal T}^{PQQ'}_{\tau_{\Lambda'} \tau_{\Lambda} \tau_{N}, \; n \, (p)} \;
\Gamma^{DDD}_{\tau_{\Lambda'} \tau_{\Lambda} \tau_{N}}(\v{k},k_{F})~.
\end{equation}
The final result obtained after the local density approximation is therefore:
\begin{eqnarray}
\label{dddisop}
\Gamma^{DDD}_{n}  & =
& 2 \{\Gamma^{DDD}_{111} + \Gamma^{DDD}_{000} + \Gamma^{DDD}_{010} + \Gamma^{DDD}_{101} \}~, \\
\Gamma^{DDD}_{p}  & =
& 2 \{5 \, \Gamma^{DDD}_{111} + \Gamma^{DDD}_{000}
- \Gamma^{DDD}_{010} - \Gamma^{DDD}_{101} \}~. \nonumber
\end{eqnarray}

Finally, we present the partial rates corresponding to the
diagram $DDDD$ of Fig.~\ref{dirgsc}. By applying the same procedure used for
$\Gamma^{DDD}_{\tau_{\Lambda'} \tau_{\Lambda} \tau_{N} }$ to Eq.~(\ref{gamppqq})
we obtain:
\begin{eqnarray}
\label{ddddgsc2}
\Gamma^{DDDD}_{\tau_{N'} \tau_{\Lambda'} \tau_{\Lambda} \tau_{N}}(\v{k},k_F)  & =
& - \, \frac{\mathcal{N}^{\, 2}(k_{F})}{(2 \pi)^2} (G_F m_{\pi}^2)^2
\left(\frac{f_{\pi}^2}{m_{\pi}^2}\right)^{2} \, \int  \,
d \v{q} \, \theta(q_0) \\
&&\times \theta(|\v{k}-\v{q}|-k_{F})
\{ (S'_{\tau_{\Lambda'}} S'_{\tau_{\Lambda}} +
P_{C, \tau_{\Lambda'}} P_{C, \tau_{\Lambda}})
{\cal V}^{2}_{C, \,\tau_{N}} \nonumber \\
&&+ (S_{\tau_{\Lambda'}}
S_{\tau_{\Lambda}}+P_{L, \tau_{\Lambda'}}  P_{L, \tau_{\Lambda}})
{\cal V}^{2}_{L, \, \tau_{N}} \nonumber \\
& &+ 2  \, ( S_{V, \tau_{\Lambda'}} S_{V, \tau_{\Lambda}} +
P_{T, \tau_{\Lambda'}} P_{T, \tau_{\Lambda}}) {\cal V}^{2}_{T, \, \tau_{N}} \} \nonumber \\
&&\times {\cal R}^{2}(-q_{0},\v{q}) {\cal I}(q_{0},\v{q})~, \nonumber
\end{eqnarray}
and
\begin{eqnarray}
\label{ddddisop}
\Gamma^{DDDD}_{n}  & =
& 4 \{\Gamma^{DDD}_{1111} + \Gamma^{DDD}_{0000} +
\Gamma^{DDD}_{0101} + \Gamma^{DDD}_{1010} \}~, \\
\Gamma^{DDDD}_{p}  & =
& 4 \{5 \, \Gamma^{DDD}_{1111} + \Gamma^{DDD}_{0000}
- \Gamma^{DDD}_{0101} - \Gamma^{DDD}_{1010} \}~, \nonumber
\end{eqnarray}
after performing the local density approximation.

In this paper the $\Gamma^{P'PQQ'}_{n\, (p)}$ exchange terms will be neglected.
Indeed, from our numerical results discussed in the next Section
it turns out that already the direct contribution
$\Gamma^{DDDD}_{n\, (p)}$ is small and approximately one order
of magnitude smaller than $\Gamma^{DDD}_{n\, (p)}$.
Moreover, according to the results obtained for the $\Gamma^{PQQ'}_{n\, (p)}$'s,
$P'PQQ'$ exchange contributions are expected to be even smaller than the
direct term $DDDD$.

\section{Results}
\label{results}
In the previous Section we have seen how the neutron-- and
proton--induced decay widths can be written in the form:
\begin{eqnarray}
\label{gamma_np}
\Gamma_{n\, (p)}&=&\Gamma^{0}_{n\, (p)}+\Gamma^{0-\rm GSC}_{n\, (p)}+
\Gamma^{\rm GSC}_{n\, (p)} \\
&\equiv & \sum_{P,Q=D,E}\Gamma^{PQ}_{n\, (p)}
+\sum_{P,Q,Q'=D,E}\Gamma^{PQQ'}_{n\, (p)}
+\sum_{P',P,Q,Q'=D,E}\Gamma^{P'PQQ'}_{n\, (p)} \nonumber~,
\end{eqnarray}
$\Gamma^{0}_{n\, (p)}$ being
the rates obtained for an uncorrelated hypernuclear ground state,
$\Gamma^{\rm GSC}_{n\, (p)}$ the rates originated by ground state
correlations and $\Gamma^{0-\rm GSC}_{n\, (p)}$
the rates resulting from the interference between uncorrelated
and correlated ground states.

For the present scheme containing
the transition amplitudes $(a)$ and $(b1)$ of Fig.~\ref{gam1gsc},
where antisymmetrization is considered for the
weak transition potential $V^{\Lambda N\to NN}$
and the nuclear residual interaction $V^{NN}$, we obtained:
two contributions to $\Gamma^{0}_{n\, (p)}$, which are
$\Gamma^{DD}_{n\, (p)}=\Gamma^{EE}_{n\, (p)}$ and
$\Gamma^{DE}_{n\, (p)}=\Gamma^{ED}_{n\, (p)}$ and
are generated by the square of amplitude $(a)$;
eight different $\Gamma^{PQQ'}_{n\, (p)}$
contributions to $\Gamma^{0-\rm GSC}_{n\, (p)}$,
which are interferences between the $(a)$ and $(b1)$ amplitudes;
sixteen different $\Gamma^{P'PQQ'}_{n\, (p)}$
contributions to $\Gamma^{\rm GSC}_{n\, (p)}$,
which originate from the square of amplitude $(b1)$.
An early evaluation of $\Gamma^0_{n\, (p)}$ has been performed
in Ref.~\cite{ba03}, while
$\Gamma^{0-\rm GSC}_{n\, (p)}$ and $\Gamma^{\rm GSC}_{n\, (p)}$
are discussed here for the first time.
Among the $\Gamma^{P'PQQ'}_{n\, (p)}$'s, here we only calculate
the direct terms $\Gamma^{DDDD}_{n\, (p)}$.

\subsection{$^{12}_\Lambda$C}

We start by discussing the relevance of the Pauli exchange terms
in $\Gamma^{0}_{n\, (p)}$ and $\Gamma^{0-\rm GSC}_{n\, (p)}$.
Our results for $\Gamma^{PQ}_{n}$ and $\Gamma^{PQ}_{p}$ are given
in Table~\ref{gamm0} for the decay of $^{12}_\Lambda$C.
Note that, for symmetry, $\Gamma^{0}_{n\, (p)}$ are twice the sum of
$\Gamma^{DD}_{n\, (p)}$ and $\Gamma^{DE}_{n\, (p)}$.
Exchange terms contribute to the uncorrelated rates for neutron--induced
(proton--induced) decays by 5.1\% (0.3\%).
Thus, they tend to increase $\Gamma_n/\Gamma_p$ while having a
very small effect on $\Gamma_1$.
\begin{table}[h]
\begin{center}
\caption{Direct and exchange $\Gamma^{PQ}_n$ and $\Gamma^{PQ}_p$ terms
for $^{12}_\Lambda$C in units of the free $\Lambda$ decay rate,
$\Gamma^0= 2.52 \cdot 10^{-6}$ eV. The first column indicates the two different
isospin channels and their sum. Note that
$\Gamma_{n\, (p)}^{DD}=\Gamma_{n\, (p)}^{EE}$ and
$\Gamma_{n\, (p)}^{DE}=\Gamma_{n\, (p)}^{ED}$.}
\label{gamm0}
\begin{tabular}{cccc}   \hline\hline
~~Channel~~ & $~~2\,\Gamma^{DD}~~$ & $~~2\,\Gamma^{DE}~~$ & $~~\Gamma^{0}~~$ \\ \hline
$\Lambda n\to nn$        & $0.146$       & $0.008$       & $0.154$            \\
$\Lambda p\to np$        & $0.469$       & $0.002$       & $0.470$            \\ \hline
sum                      & $0.615$       & $0.009$      &  $0.624$            \\
\hline\hline
\end{tabular}
\end{center}
\end{table}

In Table~\ref{gamm0gsc} we present predictions for the
$\Gamma^{PQQ'}_{n}$ and $\Gamma^{PQQ'}_{p}$ contributions derived from the
Goldstone diagrams of Fig.~\ref{antgsc}, again for $^{12}_\Lambda$C.
As expected, the direct terms $\Gamma^{DDD}_{n}$ and $\Gamma^{DDD}_{p}$
are the main contributions.
Nevertheless, the effect of antisymmetry on the two isospin channels is
significant: it increases $\Gamma^{0-\rm GSC}_{n}$ by 34\%
while decreasing $\Gamma^{0-\rm GSC}_{p}$ by 8\%.
The overall effect on $\Gamma^{0-\rm GSC}_1=
\Gamma^{0-\rm GSC}_{n}+\Gamma^{0-\rm GSC}_{p}$
is a very small increase, of 2\%.
We note that, with topologically equivalent diagrams, in Ref.~\cite{ba07b} a
similar quasi--cancellation between neutron-- and proton--induced decays
has been found in nucleon spectra calculations.
Moreover, in Ref.~\cite{ba09b} it has been shown that the
evaluation of the GSC exchange terms is important for
the rate $\Gamma_2$ as well.
We emphasize that the exact evaluation of exchange diagrams
has been mostly ignored in the literature. It is usually a quite
involved (but necessary) task, given the rapidly increasing number of terms
one has to consider when going to higher orders in the nuclear residual
interaction. Unfortunately, there is no general rule to
anticipate the need for the evaluation of exchange
terms when the corresponding direct contribution is important.
\begin{table}[h]
\begin{center}
\caption{Direct and exchange $\Gamma^{PQQ'}_{n}$
and $\Gamma^{PQQ'}_{p}$ terms for $^{12}_{\Lambda}$C obtained from
the diagrams of Fig.~\ref{antgsc}.
The first column indicates the two different
isospin channels and their sum.}
\label{gamm0gsc}
\begin{tabular}{cccccc}   \hline\hline
~~Channel~~ & $~~\Gamma^{DDD}~~$ & $~~\Gamma^{DDE}~~$ &  $~~\Gamma^{DED}~~$
& $~~\Gamma^{EDD}~~$ &  ~~~~~~~~~~~~~~~~~\\ \hline
$\Lambda n\to nn$        & $0.022$            & $-0.002$
                         & $-0.009$           & $-0.004$            &     \\
$\Lambda p\to np$        & $0.071$            & $0.005$
                         & $-0.027$           & $-0.011$            &       \\ \hline
sum                      & $0.093$            & $0.003$
                         & $-0.036$           & $-0.015$           &     \\ \hline\hline
Channel & $\Gamma^{DEE}$ & $\Gamma^{EDE}$ &  $\Gamma^{EED}$
& $\Gamma^{EEE}$ & $\Gamma^{0-\rm GSC}$ \\  \hline
$\Lambda n\to nn$        & $0.006$            & $0.008$
                         & $0.006$            & $0.002$            & $0.029$    \\
$\Lambda p\to np$        & $-0.008$           & $0.009$
                         & $0.025$            & $0.002$            & $0.066$    \\ \hline
sum                      & $-0.003$           & $0.017$
                         & $0.031$            & $0.004$            & $0.095$    \\
\hline\hline
\end{tabular}
\end{center}
\end{table}

In Table~\ref{gamm1gsc} we present the different contributions
to the rates ${\Gamma}_{n}$ and ${\Gamma}_{p}$ of Eq.~(\ref{gamma_np}).
The uncorrelated parts $\Gamma^{0}_{n}$ and $\Gamma^{0}_{p}$
dominate over the remaining ones:
$\Gamma^{0}_{1}=\Gamma^{0}_{n}+\Gamma^{0}_{p}$ constitutes the
86\% of the total ${\Gamma}_{1}$.
Then, $\Gamma^{0-\rm GSC}_{1}=\Gamma^{0-\rm GSC}_{n}+\Gamma^{0-\rm GSC}_{p}$
and $\Gamma^{\rm GSC}_{1}=\Gamma^{\rm GSC}_{n}+\Gamma^{\rm GSC}_{p}$
represent 13\% and 1\% of $\Gamma_1$, respectively.
We remind the reader that
$\Gamma^{\rm GSC}_{n\, (p)}$ are calculated from the direct
diagram $DDDD$ in Fig.~\ref{dirgsc}, while $P'PQQ'$ exchange terms
are neglected. This omission is justified
by the smallness of the direct contributions $\Gamma^{DDDD}_{n\, (p)}$:
the neglected exchange part of $\Gamma^{\rm GSC}_{1}$ should contribute
to $\Gamma_1$ by less than 1\%. Thus, a challenging calculation of the
fifteen $P'PQQ'$ exchange diagrams can be reasonably avoided.
\begin{table}[h]
\begin{center}
\caption{Predictions for the one--nucleon induced decay rates
of Eq.~(\ref{gamma_np}) for $^{12}_\Lambda$C. The first column indicates the two different
isospin channels and their sum.}
\label{gamm1gsc}
\begin{tabular}{ccccc}   \hline\hline
~~Channel~~ & $~~\Gamma^{0}~~$ & $~~\Gamma^{0-\rm GSC}~~$ &
$~~\Gamma^{\rm GSC}~~$ & $~~\Gamma~~$ \\  \hline
$\Lambda n\to nn$ & $0.154$  & $0.029$ & $0.002$ & $0.185$ \\
$\Lambda p\to np$ & $0.470$  & $0.066$ & $0.008$ & $0.544$  \\ \hline
sum               & $0.624$  & $0.095$ & $0.010$ & $0.729$  \\ \hline\hline
\end{tabular}
\end{center}
\end{table}

Our predictions for the one-- and two--nucleon induced decay rates
for $^{12}_\Lambda$C are given in Table~\ref{gamm12} and compared with
the most recent data by KEK~\cite{Ki09} and FINUDA~\cite{Ag09}.
For completeness,
we report results without and with the inclusion of antisymmetrization and GSC.
It should be noted that the hypernuclear ground state normalization
function $\mathcal{N}(k_{F})$ of Eq.~(\ref{norconst})
equally affects ${\Gamma}_{1}$ and ${\Gamma}_{2}$. This function
is not identically equal to one only when GSC are present. Therefore,
the $\Gamma_{1}$ result without GSC and with exchange terms of Table~\ref{gamm12}, 0.74,
is bigger than the prediction for $\Gamma^{0}_1$ of Table~\ref{gamm1gsc}, 0.62,
which has been obtained instead by including both GSC
and antisymmetrization in the normalization function.
This comparison gives an idea of the importance of a proper
normalization of the hypernuclear ground state.
GSC produces a sizable increase in the value of $\Gamma_{\rm NM}$, thanks
to the opening of the two--nucleon induced channel, while
$\Gamma_1$ remains practically unaffected.
The effect of GSC on the ${\Gamma}_{n}/{\Gamma}_{p}$ ratio is
a small increase of 4\%, which is due entirely
to the exchange terms in $\Gamma^{0-\rm GSC}_n$
and $\Gamma^{0-\rm GSC}_p$ (see Table~\ref{gamm0gsc}).
Antisymmetrization on the contrary introduce an increase
of $\Gamma_1$ and a reduction of $\Gamma_2$, and as a result
a sizable reduction of $\Gamma_2/\Gamma_1$.
We conclude that GSC are important to get agreement with data
on $\Gamma_{\rm NM}$, while antisymmetrization is
crucial to reproduce the data for $\Gamma_2/\Gamma_1$.
Note indeed that only with the set of results including both exchange terms
and GSC we can achieve an overall agreement with all data.
\begin{table}[h]
\begin{center}
\caption{The non--mesonic weak decay widths of $^{12}_{\Lambda}$C.
Results are given without and with the contributions of
antisymmetrization and ground state correlations.
The most recent data, from KEK~\cite{Ki09} and FINUDA~\cite{Ag09},
are given for comparison.}
\label{gamm12}
\resizebox*{\textwidth}{!}{
\begin{tabular}{cccccccc}   \hline\hline
Ant./GSC & $\Gamma_n$ & $\Gamma_p$ &
${\Gamma}_{1}$ & ${\Gamma}_{2}$ & $\Gamma_{\rm NM}$
 & $\Gamma_n/\Gamma_p$ & ${\Gamma}_{2}/{\Gamma}_{\rm NM}$ \\  \hline
no/no   & 0.15 & 0.47 & 0.62 &   0  & 0.62 & 0.31 &  0\\
yes/no  & 0.18 &       0.56 &  0.74 &   0  & 0.74 &  0.33 &  0\\
no/yes  &  0.15 & 0.47 & 0.61 &  0.31 & 0.91 & 0.31 & 0.50 \\
yes/yes &  0.19 & 0.55 & 0.73 &  0.25 & 0.98 & 0.34 & 0.26
\\ \hline
KEK& $0.23\pm 0.08$ & $0.45\pm 0.10$ &
$0.68\pm 0.13$ & $0.27\pm 0.13$ &
$0.95 \pm 0.04$ &  $0.51 \pm 0.13 \pm 0.05$ & $0.29\pm 0.13$ \\
FINUDA & & & & & & & $0.24\pm 0.10$ \\ \hline\hline
\end{tabular}}
\end{center}
\end{table}

Despite this agreement, we have
to admit that more refined and systematic theoretical studies should be performed before one can
reach definite conclusions from the comparison between theory and experiment. For instance, the
result obtained for ${\Gamma}_{\rm NM}$ requires a comment on the eventual inclusion of the full
set of diagrams stemming from the amplitudes in Figs.~\ref{gam1gsc} and~\ref{gam2gsc} and
eventually from other amplitudes. At first glance, one may think that the final outcome from all
these diagrams would be a bigger value for ${\Gamma}_{\rm NM}$, thus spoiling the good
agreement with data of the present result.
This is not necessarily the case, for two reasons.
First, the amplitudes $(d1)$ and $(d2)$ in Fig.~\ref{gam1gsc} and the amplitude
$(c)$ in Fig.~\ref{gam2gsc} originate from $1\Delta1p2h$ GSC. The inclusion of these correlation
amplitudes requires the introduction of new terms in the ground state normalization function
(\ref{norconst}); this leads to a reduction of the individual values for each decay width,
including the ones we have obtained above. From the previous studies in
Refs.~\cite{ba09,ba09b} one observes the following property, introduced by ground state
normalization: a certain redistribution of the total non--mesonic decay strength among the
partial contributions occurs when new self--energy terms are included. Secondly, the presence of
several additional self--energy diagrams which are interference terms between amplitudes could
also bring to a reduction of the decay rates $\Gamma_1$ and $\Gamma_2$.

\subsection{Medium and heavy hypernuclei}
\begin{figure}[htb]
\begin{center}
\mbox{\epsfig{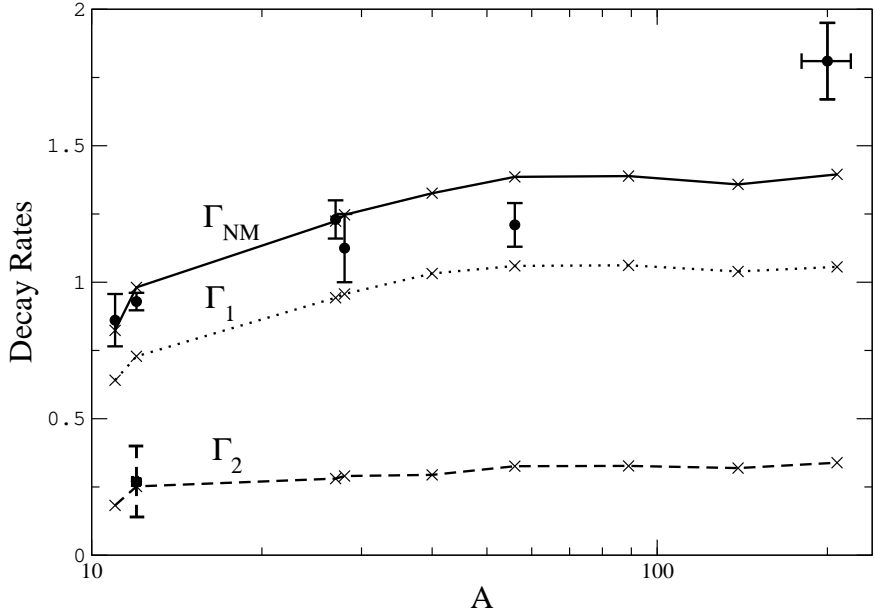}}
\caption{The predictions for the decay rates $\Gamma_1$, $\Gamma_2$ and
$\Gamma_{\rm NM}=\Gamma_1+\Gamma_2$ are given
as a function of the hypernuclear mass number $A$. The
results for $\Gamma_{\rm NM}$ are compared with
experimental data for $^{11}_\Lambda$B~\cite{Sa05},
$^{12}_\Lambda$C~\cite{Ou08}, $^{27}_\Lambda$Al~\cite{Sa05},
$^{28}_\Lambda$Si~\cite{Sa05}, $^{56}_\Lambda$Fe~\cite{Sa05}
and for the region of $A$ between $180$ and $220$ \cite{Cosy}.
The datum for $\Gamma_2$ is from Ref.~\cite{Ki09}.}
\label{fm-to-h}
\end{center}
\end{figure}

In order to have a further indication of the reliability of our framework,
which adopts the local density approximation to obtain results for finite hypernuclei,
we have extended the calculation to medium and heavy $\Lambda$ hypernuclei.
All the GSC contributions and the antisymmetrization terms
discussed in detail for $^{12}_\Lambda$C have been taken into account.
The results we have obtained are given in Table~\ref{m-to-h} and are compared with
recent data in Figure~\ref{fm-to-h}.

The GSC--free rate $\Gamma_1^0$ represents 86\% of the
rate $\Gamma_1=\Gamma_1^0+\Gamma_1^{0-\rm GSC}+\Gamma_1^{\rm GSC}$
for $^{12}_\Lambda$C. For increasing hypernuclear mass number $A$,
this contribution decreases and reaches 81\% for $^{208}_\Lambda$Pb.
As expected, GSC contributions are thus more important for
heavy hypernuclei.

The one-- and two--nucleon induced rates increase with $A$ and rapidly saturate.
Saturation is expected to begin for those hypernuclei whose radius
becomes sensitively larger than the range of the non--mesonic
processes. The fact that for $^{40}_\Lambda$Ca and $^{208}_\Lambda$Pb
we obtain very similar predictions informs us that in $^{208}_\Lambda$Pb
the non--mesonic decay (both one-- and two--nucleon stimulated)
involve the same nucleon shells which participate in the decay of $^{40}_\Lambda$Ca.
Indeed, the $\Lambda$ wave function ($s$ level of the
$\Lambda$--nucleus mean potential) is well overlapped to the
hypernuclear core already in $^{40}_\Lambda$Ca.

It should be noted that the slight decrease of the non--mesonic rate
$\Gamma_{\rm NM}$ going from $^{89}_\Lambda$Y to $^{139}_\Lambda$La
is due to the special value of the oscillator parameter $\hbar \omega$ adopted
for this hypernucleus. Such a parameter, which is obtained as the difference
between the measured $s$ and $p$ $\Lambda$ energy levels in $^{139}_\Lambda$La,
is indeed smaller than the values measured for the two neighboring hypernuclei
of our calculation, $^{89}_\Lambda$Y and $^{208}_\Lambda$Pb.

The contribution of the two--nucleon induced width
is almost independent of the hypernuclear mass number and oscillates
between 22 and 26\% of $\Gamma_{\rm NM}$.
We note from Figure~\ref{fm-to-h} that the datum recently determined at KEK,
$\Gamma_2=0.27\pm 0.13$ \cite{Ki09}, is well reproduced by our calculation.
Also the recent determination obtained by FINUDA \cite{Ag09} of
$\Gamma_2/\Gamma_{\rm NM}=0.22\pm 0.08$ for hypernuclei from $^5_\Lambda$He
to $^{16}_\Lambda$O is in agreement with our predictions.

Concerning $\Gamma_{\rm NM}$, the agreement of our predictions with data is
also rather good.
The only exception is the large underestimation of the datum for the
$A\simeq 200$ region, which however is also difficult to reconcile with the
decay rate measured at KEK for $^{56}_\Lambda$Fe. No known mechanism
can be responsible for a large increase in the non--mesonic decay rate when
going from $^{56}_\Lambda$Fe to the $A\simeq 200$ region.
Concerning the datum for $A\simeq 200$, we have to note that,
given the difficulty in employing direct timing methods
for heavy hypernuclei, it has been obtained in experiments
(performed at COSY, Juelich \cite{Cosy}) which measured the fission fragments
(which are supposed to be generated by the non--mesonic decay)
emitted by hypernuclei produced in proton--nucleus reactions.
Large uncertainties affect such delayed fission experiments, because
of the limited precision of the employed recoil shadow method.
The produced hypernuclei cannot be unambiguously identified
with this method. It is also possible that mechanisms other than the
non--mesonic decay contributed to hypernuclear fission in these experiments.
The datum reported in Figure~\ref{fm-to-h} has been obtained as an
average from measurements for hypernuclei produced in proton--Au,
proton--Bi and proton--U reactions.

\begin{table}[h]
\begin{center}
\caption{Decay rates predicted for medium to heavy hypernuclei.}
\begin{tabular}{ccccc}   \hline\hline
${\rm Hypernucleus}$ & $~~~~~\Gamma^0_1~~~~~$ &
$~~~~~\Gamma_1~~~~~$ & $~~~~~\Gamma_2~~~~~$
& $~~~~~\Gamma_{\rm NM}~~~~~$ \\  \hline
$^{11}_\Lambda$B   & 0.56 & 0.64 & 0.18 & 0.82   \\
$^{12}_\Lambda$C   & 0.62 & 0.73 & 0.25 & 0.98   \\
$^{27}_\Lambda$Al  & 0.80 & 0.94 & 0.28 & 1.22   \\
$^{28}_\Lambda$Si  & 0.81 & 0.96 & 0.29 & 1.25   \\
$^{40}_\Lambda$Ca  & 0.87 & 1.03 & 0.29 & 1.33     \\
$^{56}_\Lambda$Fe  & 0.88 & 1.06 & 0.33 & 1.39   \\
$^{89}_\Lambda$Y   & 0.87 & 1.06 & 0.33 & 1.39   \\
$^{139}_\Lambda$La & 0.86 & 1.04 & 0.32 & 1.36     \\
$^{208}_\Lambda$Pb & 0.86 & 1.06 & 0.34 & 1.40   \\
\hline\hline
\label{m-to-h}
\end{tabular}
\end{center}
\end{table}

We think that the results of the evaluation for medium and heavy hypernuclei
are encouraging: they give us some confidence
in using the local density approximation for obtaining results in finite
hypernuclei, even in light systems such as $^{12}_\Lambda$C.

\subsection{Closing remarks}

Before concluding, we make here some further comments on our calculation.
Through our work
we wish to emphasize the importance of a detailed many--body treatment of
non--mesonic decay.
This requires the identification and evaluations of a large number of
diagrams, working on a step--by--step basis
with the perspective of reaching the condition in which the terms
that are not taken into account can be safely neglected.
Considering the evolution in the predictions obtained in recent works
(see especially Refs.~\cite{ba09,ba09b})
and here, this stability of results has not been achieved yet,
and new many--body terms must be considered.
In our opinion, one should explore the dependencies of predictions on the
weak transition potential model only after these complicated
many--body aspects are properly understood.
Finally, one should attempt to reach a detailed
agreement with experiment for $\Gamma_{\rm NM}$, $\Gamma_n/\Gamma_p$
and $\Gamma_2/\Gamma_{\rm NM}$
and thus extract sensible information on strangeness--changing
baryon interactions. From the experimental side,
new and improved data are expected from FINUDA@Daphne~\cite{FI},
J–PARC~\cite{jparc,jparc-exp} and GSI~\cite{GSI}.
A direct experimental identification of the two--nucleon induced channels
together with the measurement of $\Gamma_2$ is a question of particular
importance.

We end this Section with a comment to emphasize the importance of
evaluating exchange terms. In our many--body inspired calculation,
such terms are considered together with GSC contributions,
which are included on the same ground for one-- and two--nucleon induced decays.
GSC and exchange terms improve
by 10\% the value of ${\Gamma}_{n}/{\Gamma}_{p}$.
Once GSC are included, antisymmetrization turns out to be
particularly important for both the one-- and the two--nucleon induced channels,
reducing ${\Gamma}_{2}$ by 18\% and increasing ${\Gamma}_{1}$ by 20\%.
It would thus be pointless to neglect exchange terms and
evaluate only direct ones. Although the introduction
of antisymmetry is a difficult task in a many--body framework, one
should evaluate all those exchange diagrams which
are companions of a direct diagram which one knows to be relevant.

\section{Conclusions}
\label{conclusions}

In this contribution we have studied the
effects of GSC in the non--mesonic weak decay of $\Lambda$ hypernuclei.
A non--relativistic nuclear matter scheme has been adopted together with
the local density approximation, for calculations in hypernuclei
ranging from $^{11}_\Lambda$B to $^{208}_\Lambda$Pb.
All isospin channels contributing to one--
and two--nucleon induced decays have been considered.
The employed weak transition potential contains the exchange of mesons of the
pseudoscalar and vector octets, $\pi$, $\eta$, $K$, $\rho$, $\omega$ and $K^*$.
The residual strong interaction, responsible for GSC, has been modeled
on a Bonn potential based on $\pi$--, $\rho$--, $\sigma$-- and $\omega$--exchange.

By using the Goldstone diagrams technique,
GSC have been introduced on the same footing for one-- and two--nucleon
stimulated decays. The normalization of the hypernuclear ground state
introduced by GSC has been taken into account.
We have devoted particular attention to those
GSC affecting the decay widths $\Gamma_{n}$ and $\Gamma_{p}$.
The many--body $\Lambda$ self--energy terms we have considered are originated by
the transition amplitudes $(a)$ and $(b1)$ of Fig.~\ref{gam1gsc}
(for one--nucleon induced decays)
and by the amplitude $(a)$ of Fig.~\ref{gam2gsc}
(for two--nucleon induced decays). Our approach
embodies fermion antisymmetry, i.e., both direct and exchange interactions
are considered in the various diagrams. Concerning one--nucleon induced decays,
we have evaluated GSC--free rates $\Gamma^0_{n\, (p)}$,
generated by amplitude $(a)$, purely GSC terms $\Gamma^{\rm GSC}_{n\, (p)}$,
produced by amplitude $(b1)$, and interference terms $\Gamma^{0-\rm GSC}_{n\, (p)}$
between uncorrelated and correlated hypernuclear ground states, i.e.,
between amplitudes $(a)$ and $(b1)$.

The dominant contribution to $\Gamma_1=\Gamma^0_{1}+\Gamma^{0-\rm GSC}_{1}
+\Gamma^{\rm GSC}_{1}$ turned out to be $\Gamma^0_{1}=\Gamma^0_{n}+\Gamma^0_{p}$.
For $^{12}_\Lambda$C,
$\Gamma^{0-\rm GSC}_{1}=\Gamma^{0-\rm GSC}_{n}+\Gamma^{0-\rm GSC}_{p}$
and $\Gamma^{\rm GSC}_{1}=\Gamma^{\rm GSC}_{n}+\Gamma^{\rm GSC}_{p}$
represented 13\% and 1\% of the rate $\Gamma_1$, respectively;
GSC are thus responsible for 14\% of the one--nucleon induced width
(such contribution increases up to 19\% for $^{208}_\Lambda$Pb).
The above results justify the fact that
we have neglected the exchange terms in $\Gamma^{\rm GSC}_{n\, (p)}$.
Exchange contributions are rather relevant in the calculation of
$\Gamma^{0-\rm GSC}_{n\, (p)}$ (for $^{12}_\Lambda$C,
they increase $\Gamma^{0-\rm GSC}_n$ by 34\%
and decreases $\Gamma^{0-\rm GSC}_p$ by 8\%), while only
scarcely contribute to $\Gamma^{0}_{n\, (p)}$.
GSC and exchange terms together increase the value of
${\Gamma}_{n}/{\Gamma}_{p}$ for $^{12}_\Lambda$C by 10\%.
Thanks to the opening of the two--nucleon induced channel,
GSC produces a sizable increase (of 32\% for $^{12}_\Lambda$C
when exchange terms are included) in the value of
$\Gamma_{\rm NM}=\Gamma_1+\Gamma_2$.

The agreement among our final results
and recent data is quite good and clearly demonstrates the necessity
of including GSC and antisymmetrization effects.
Nevertheless, we believe that a refinement of the present scheme must be pursued.
Additional many--body terms should be considered,
involving for instance the $\Delta(1232)$ resonance.
Only after a certain stability of predictions is reached
within such a microscopic approach one should
explore the dependencies on the weak transition potential model
and determine, through detailed comparison with experiment,
sensible information on strangeness--changing baryon interactions.

\section*{Acknowledgments}
This work has been partially supported by the CONICET,
under contract PIP 6159. This Research is part of the EU Initiative
FP7-Project HadronPhysics2 under Project number 227431.
We would like to thank A. Ramos
and F. Krmpotic for the helpful discussion and the careful reading
of the manuscript.

\section*{Appendix A}
\label{APPENDB}
In this Appendix we present explicit expressions for the decay rates
$\Gamma^{PQQ'}_{n\, (p)}$ with $PQQ' \neq DDD$
associated to the Goldstone diagrams of Fig.~\ref{antgsc}
and contributing to Eq.~(\ref{rpa2}). In the main text, these widths
have been written as:
\begin{equation}
\label{iso-sum}
\Gamma^{PQQ'}_{n\, (p)}=
\sum_{\tau_{\Lambda'}, \tau_{\Lambda}, \tau_{N}=0,1}
{\cal T}^{PQQ'}_{\tau_{\Lambda'} \tau_{\Lambda} \tau_{N}, \; n \, (p)} \;
\Gamma^{PQQ'}_{\tau_{\Lambda'} \tau_{\Lambda} \tau_{N}}(\v{k},k_{F})~,
\end{equation}
where
\begin{eqnarray}
\label{appen-gamma}
\Gamma^{PQQ'}_{\tau_{\Lambda'} \tau_{\Lambda} \tau_{N}}(\v{k},k_F) & =
& \mathcal{N}^{\, 2}(k_{F})
\frac{1}{4} \frac{(-1)^{n}}{(2 \pi)^8} (G_F m_{\pi}^2)^2
\frac{f_{\pi}^2}{m_{\pi}^2} \\
&&\times \int  \int \int \, d \v{q} \, d \v{h} \, d \v{h}' \;
{\cal W}^{PQQ'}_{\tau_{\Lambda'} \tau_{\Lambda} \tau_{N}}(q, q', t) \nonumber \\
&&\times \Theta(k,q,q',t,h,h',k_{F}) \frac{1}{- \varepsilon^{PQQ'}_{2p2h}} \;
\delta(q_0 - (\varepsilon_{\v{h}'+\v{q}}-\varepsilon_{\v{h}'}))~. \nonumber
\end{eqnarray}
The isospin index $\tau_{\Lambda}$ ($\tau_{\Lambda'}$)
of the weak transition potential
is associated to an energy--momentum $q$ ($q'$), while the nuclear strong interaction
isospin index is $\tau_{N}$ and the corresponding energy--momentum $t$.
In the following subsections we give the functions
${\cal W}^{PQQ'}_{\tau_{\Lambda'} \tau_{\Lambda} \tau_{N}}(q, q', t)$ and
$\Theta(k,q,q',t,h,h',k_{F})$,
the energy denominator $\varepsilon^{PQQ'}_{2p2h}$ and $n$
(the number of crossing between fermionic lines) for the various
cases. Finally, we show the isospin sums of Eq.~(\ref{iso-sum}).

\subsection*{i) $\bf \Gamma^{DDE}_{n\, (p)}$}
The ${\cal W}^{DDE}_{\tau_{\Lambda'} \tau_{\Lambda} \tau_{N}}(q, q', t)$
function, where $q'=q$ and $t=h'-h+q$, is identical to the
${\cal S}^{ded}_{\tau' \tau_{N} \tau}(q, q', t)$ function in
Eq.~(A.1) of Ref.~\cite{ba07b}.
Moreover:
\begin{eqnarray}
\label{thetaDDE}
\Theta(k,q,q',t,h,h',k_{F}) & = &
\theta(q_0) \theta(|\v{k}-\v{q}|-k_{F}) \theta(|\v{q}-\v{h}|-k_{F}|) \\
&& \times \theta(k_{F}-|\v{h}|)\theta(|\v{q}+\v{h}'|-k_{F}|)
\theta(k_{F}-|\v{h}'|)~, \nonumber
\end{eqnarray}
\begin{equation}
\varepsilon^{DDE}_{2p2h}=\varepsilon^{DDD}_{2p2h}
\equiv k_0-\varepsilon_{\v{k}-\v{q}}+\varepsilon_{\v{h}-\v{q}}-
\varepsilon_{\v{h}}-V_N~,
\end{equation}
and $n=0$. The isospin sums are given by:
\begin{eqnarray}
\label{decDDE}
\Gamma^{DDE}_{n}& = & -\Gamma^{DDE}_{111}+\Gamma^{DDE}_{000}+3 \Gamma^{DDE}_{101}
+\Gamma^{DDE}_{110}-\Gamma^{DDE}_{011} +\Gamma^{DDE}_{100} \nonumber \\
&&+3 \Gamma^{DDE}_{001}+\Gamma^{DDE}_{010}~, \nonumber \\
\Gamma^{DDE}_{p}& = & -5 \Gamma^{DDE}_{111}+\Gamma^{DDE}_{000}-3 \Gamma^{DDE}_{101}
+5 \Gamma^{DDE}_{110}+\Gamma^{DDE}_{011} -\Gamma^{DDE}_{100} \nonumber \\
&&+3 \Gamma^{DDE}_{001}-\Gamma^{DDE}_{010}~. \nonumber
\end{eqnarray}

\subsection*{ii) $\bf \Gamma^{DED}_{n\, (p)}$}
The ${\cal W}^{DED}_{\tau_{\Lambda'} \tau_{\Lambda} \tau_{N}}(q, q', t)$
function, where $q'=k-h$ and $t=q$, is identical to the
${\cal S}^{dde}_{\tau' \tau_{N} \tau}(q, q', t)$ function in
Eq.~(A.3) of Ref.~\cite{ba07b}.
Moreover:
\begin{eqnarray}
\label{thetaDED}
\Theta(k,q,q',t,h,h',k_{F}) & = &
\theta(q_0) \theta(|\v{k}-\v{q}|-k_{F}) \theta(|\v{q}-\v{h}|-k_{F}|) \\
&& \times \theta(k_{F}-|\v{h}|)
\theta(|\v{q}+\v{h}'|-k_{F}|)
\theta(k_{F}-|\v{h}'|)~, \nonumber
\end{eqnarray}
\begin{equation}
\varepsilon^{DED}_{2p2h}=\varepsilon^{DDD}_{2p2h}~,
\end{equation}
and $n=0$. The isospin sums are given by:
\begin{eqnarray}
\label{decDED}
\Gamma^{DED}_{n}& = & -\Gamma^{DED}_{111}+\Gamma^{DED}_{000}+\Gamma^{DED}_{101}
+3 \Gamma^{DED}_{110}-\Gamma^{DED}_{011}+\Gamma^{DED}_{100} \nonumber \\
&&+3 \Gamma^{DED}_{001}+3\Gamma^{DED}_{010}~, \nonumber \\
\Gamma^{DED}_{p}& = & -5 \Gamma^{DED}_{111}+\Gamma^{DED}_{000}+5 \Gamma^{DED}_{101}
-3 \Gamma^{DED}_{110}+\Gamma^{DED}_{011}-\Gamma^{DED}_{100} \nonumber \\
&&- \Gamma^{DED}_{001}+3\Gamma^{DED}_{010}~. \nonumber
\end{eqnarray}

\subsection*{iii) $\bf \Gamma^{EDD}_{n(p)}$}
The ${\cal W}^{EDD}_{\tau_{\Lambda'} \tau_{\Lambda} \tau_{N}}(q, q', t)$
function, where $q'=k-q-h'$ and $t=q$, is identical to the
${\cal S}^{dde}_{\tau' \tau_{N} \tau}(q, q', t)$ function in
Eq.~(A.3) of Ref.~\cite{ba07b}.
Moreover:
\begin{eqnarray}
\label{thetaEDD}
\Theta(k,q,q',t,h,h',k_{F}) & = &
\theta(q_0) \theta(|\v{k}-\v{q}|-k_{F}) \theta(|\v{q}-\v{h}|-k_{F}|) \\
&& \times \theta(k_{F}-|\v{h}|) \theta(|\v{q}+\v{h}'|-k_{F}|)
\theta(k_{F}-|\v{h}'|)~, \nonumber
\end{eqnarray}
\begin{equation}
\varepsilon^{EDD}_{2p2h}=\varepsilon^{DDD}_{2p2h}~,
\end{equation}
and $n=1$. The isospin sums are given by:
\begin{eqnarray}
\label{decEDD}
\Gamma^{EDD}_{n}& = & -\Gamma^{EDD}_{111}+\Gamma^{EDD}_{000}- \Gamma^{EDD}_{101}
+ 3 \Gamma^{EDD}_{110}+\Gamma^{EDD}_{011} +3 \Gamma^{EDD}_{100} \nonumber \\
&&+\Gamma^{EDD}_{001}+\Gamma^{EDD}_{010}~, \nonumber \\
\Gamma^{EDD}_{p}& = & -5 \Gamma^{EDD}_{111}+\Gamma^{EDD}_{000}+\Gamma^{EDD}_{101}
- 3 \Gamma^{EDD}_{110}+ 5 \Gamma^{EDD}_{011}+ 3 \Gamma^{EDD}_{100} \nonumber \\
&& - \Gamma^{EDD}_{001}-\Gamma^{EDD}_{010}~. \nonumber
\end{eqnarray}

\subsection*{iv) $\bf \Gamma^{DEE}_{n(p)}$}
The ${\cal W}^{DEE}_{\tau_{\Lambda'} \tau_{\Lambda} \tau_{N}}(q, q', t)$
function, where $q'=k-h$ and $t=h-h'-q$, is identical to the
${\cal S}^{eed}_{\tau' \tau_{N} \tau}(q, q', t)$ function in
Eq.~(A.7) of Ref.~\cite{ba07b}.
Moreover:
\begin{eqnarray}
\label{thetaDEE}
\Theta(k,q,q',t,h,h',k_{F}) & = &
\theta(q_0) \theta(|\v{k}-\v{q}|-k_{F}) \theta(|\v{q}-\v{h}|-k_{F}|) \\
& & \times \theta(k_{F}-|\v{h}|)\theta(|\v{q}+\v{h}'|-k_{F}|)
\theta(k_{F}-|\v{h}'|)~, \nonumber
\end{eqnarray}
\begin{equation}
\varepsilon^{DEE}_{2p2h}=\varepsilon^{DDD}_{2p2h}~,
\end{equation}
and $n=1$. The isospin sums are given by:
\begin{eqnarray}
\label{decdee}
\Gamma^{DEE}_{n}& = & 5 \Gamma^{DEE}_{111}+\Gamma^{DEE}_{000}+ \Gamma^{DEE}_{101}
+\Gamma^{DEE}_{110}+5\Gamma^{DEE}_{011} +\Gamma^{DEE}_{100} \nonumber \\
&& + \Gamma^{DEE}_{001}+\Gamma^{DEE}_{010}~, \nonumber \\
\Gamma^{DEE}_{p}& = & -2 \Gamma^{DEE}_{111}-4 \Gamma^{DEE}_{101}
+4 \Gamma^{DEE}_{110}+2\Gamma^{DEE}_{011} +2\Gamma^{DEE}_{100} \nonumber\\
&&+2 \Gamma^{DEE}_{001}+2\Gamma^{DEE}_{010}~. \nonumber
\end{eqnarray}

\subsection*{v) $\bf \Gamma^{EDE}_{n(p)}$}
The ${\cal W}^{EDE}_{\tau_{\Lambda'} \tau_{\Lambda} \tau_{N}}(q, q', t)$
function, where $q'=k-q-h'$ and $t=h'-h+q$, is identical to the
${\cal S}^{eed}_{\tau' \tau_{N} \tau}(q, q', t)$ function in
Eq.~(A.7) of Ref.~\cite{ba07b}.
Moreover:
\begin{eqnarray}
\label{thetaEDE}
\Theta(k,q,q',t,h,h',k_{F}) & = &
\theta(q_0) \theta(|\v{k}-\v{q}|-k_{F}) \theta(|\v{q}-\v{h}|-k_{F}|) \\
& & \times \theta(k_{F}-|\v{h}|) \theta(|\v{q}+\v{h}'|-k_{F}|)
\theta(k_{F}-|\v{h}'|)~, \nonumber
\end{eqnarray}
\begin{equation}
\varepsilon^{EDE}_{2p2h}=\varepsilon^{DDD}_{2p2h}~,
\end{equation}
and $n=1$. The isospin sums are given by:
\begin{eqnarray}
\label{decEDE}
\Gamma^{EDE}_{n}& = & -\Gamma^{EDE}_{111}+\Gamma^{EDE}_{000}+3 \Gamma^{EDE}_{101}
+\Gamma^{EDE}_{110}-\Gamma^{EDE}_{011} +\Gamma^{EDE}_{100} \nonumber\\
&&+ 3 \Gamma^{EDE}_{001}+\Gamma^{EDE}_{010} \nonumber \\
\Gamma^{EDE}_{p}& = & 4 \Gamma^{EDE}_{111}+6 \Gamma^{EDE}_{101}
-4 \Gamma^{EDE}_{110}-2\Gamma^{EDE}_{011}
+2\Gamma^{EDE}_{100}+2\Gamma^{EDE}_{010} \nonumber
\end{eqnarray}

\subsection*{vi) $\bf \Gamma^{EED}_{n(p)}$}
The ${\cal W}^{EED}_{\tau_{\Lambda'} \tau_{\Lambda} \tau_{N}}(q, q', t)$
function, where $q'=k-h$ and $t=k-q-h'$, is identical to the
${\cal S}^{ede}_{\tau' \tau_{N} \tau}(q, q', t)$ function in
Eq.~(A.5) of Ref.~\cite{ba07b}.
Moreover:
\begin{eqnarray}
\label{thetaEED}
\Theta(k,q,q',t,h,h',k_{F}) & = &
\theta(q_0) \theta(|\v{k}-\v{q}|-k_{F}) \theta(|\v{q}+\v{h}+\v{h}'-\v{k}|-k_{F}|)\\
& & \times \theta(k_{F}-|\v{h}|) \theta(|\v{q}+\v{h}'|-k_{F}|)
\theta(k_{F}-|\v{h}'|)~, \nonumber
\end{eqnarray}
\begin{equation}
\varepsilon^{EED}_{2p2h}=k_0-\varepsilon_{\v{h}}+
\varepsilon_{\v{q}+\v{h}+\v{h}'-\v{k}}-\varepsilon_{\v{q}+\v{h'}}-V_N~,
\end{equation}
and $n=1$. The isospin sum are given by:
\begin{eqnarray}
\label{decEED}
\Gamma^{EED}_{n}& = & -\Gamma^{EED}_{111}+\Gamma^{EED}_{000}+ \Gamma^{EED}_{101}
+3\Gamma^{EED}_{110}-\Gamma^{EED}_{011}
+\Gamma^{EED}_{100}~, \nonumber\\
&&+\Gamma^{EED}_{001}+3\Gamma^{EED}_{010} \nonumber \\
\Gamma^{EED}_{p}& = & 4 \Gamma^{EED}_{111}-4 \Gamma^{EED}_{101}
+6 \Gamma^{EED}_{110}-2\Gamma^{EED}_{011}
+2\Gamma^{EED}_{100}+2 \Gamma^{EED}_{001}~. \nonumber
\end{eqnarray}

\subsection*{vii) $\bf \Gamma^{EEE}_{n(p)}$}
The ${\cal W}^{EEE}_{\tau_{\Lambda'} \tau_{\Lambda} \tau_{N}}(q, q', t)$
function, where $q'=k-h$ and $t=h+q-k$, is identical to the
${\cal S}^{eee}_{\tau' \tau_{N} \tau}(q, q', t)$ function in
Eq.~(A.9) of Ref.~\cite{ba07b}.
Moreover:
\begin{eqnarray}
\label{thetaEEE}
\Theta(k,q,q',t,h,h',k_{F}) & = &
\theta(q_0) \theta(|\v{k}-\v{q}|-k_{F}) \theta(|\v{h}'+\v{q}|-k_{F}|)\\
& & \times \theta(k_{F}-|\v{h}'|)\theta(k_{F}-|\v{k}-\v{h}+\v{h}'|)
\theta(k_{F}-|\v{h}|)~, \nonumber
\end{eqnarray}
\begin{equation}
\varepsilon^{EEE}_{2p2h}=k_0-\varepsilon_{\v{h}}
+\varepsilon_{\v{h}'}-\varepsilon_{\v{k}-\v{h}+\v{h}'}-V_N~,
\end{equation}
and $n=0$. The isospin sums are given by:
\begin{eqnarray}
\label{deceee}
\Gamma^{EEE}_{n}& = & 5\Gamma^{EEE}_{111}+\Gamma^{EEE}_{000}+\Gamma^{EEE}_{101}
+\Gamma^{EEE}_{110}+5\Gamma^{EEE}_{011}
+\Gamma^{EEE}_{100} \nonumber \\
&&+ \Gamma^{EEE}_{001}+\Gamma^{EEE}_{010}~, \nonumber \\
\Gamma^{EEE}_{p}& = & 7 \Gamma^{EEE}_{111}+\Gamma^{EEE}_{000}+5 \Gamma^{EEE}_{101}
+5 \Gamma^{EEE}_{110}+\Gamma^{EEE}_{011}-\Gamma^{EEE}_{100} \nonumber \\
&&- \Gamma^{EEE}_{001}-\Gamma^{EEE}_{010}~. \nonumber
\end{eqnarray}

\section*{Appendix B}
\label{APPENDa}
The explicit expressions of the function ${\cal R}(q_{0},\v{q})$
of Eq.~(\ref{real}) reads:
\begin{eqnarray}
\label{real2}
{\cal R}(q_{0},\v{q}) & = & \frac{\pi}{(2 \pi)^{3}} \, \frac{m}{q} \,
\left\{ \frac{m^{2}}{q^{2}} \, \left[2\left(q_0-\frac{q^{2}}{2 m}\right) \frac{q}{m} k_F
+\left(\left(q_0 - \frac{q^{2}}{2 m}\right)^{2}-\frac{q^{2}}{m^{2}} k^{2}_F\right)
\right. \right. \nonumber \\
&&
\times \left. \, \ln \left|\frac{2 m q_0-q^{2}-2 q k_F}
{2 m q_0- q^{2} +2 q k_F}\right|\right] + \theta(2 k_F-q)
\, \left[-\frac{m^{2}}{q^{2}} \,\left(2 q_0  \frac{q}{m}
\left(k_F- \frac{q}{2}\right) \right. \right.
\nonumber \\
&&
\left. +\left(q^{2}_0 - \frac{q^{2}}{m^{2}}\left(k_F- \frac{q}{2}\right)^{2}\right)
\ln \left|\frac{2 m q_0+q^{2}-2 q k_F}
{2 m q_0- q^{2} +2 q k_F}\right|\right) + q \left(\frac{q}{4}-k_F\right)
\nonumber \\
&& \left. \left. \times \ln \left|\frac{2 m q_0-q^{2}+2 q k_F}
{2 m q_0+ q^{2} -2 q k_F}\right|
-q_0 m \ln \left|\frac{q^{2}_0 m^{2}-q^{2}(k_F-q/2)^{2}}
{m^{2} q^{2}_0}\right| \right] \right\}~, \nonumber
\end{eqnarray}
where $q=|\v{q}|$ and $m$ is the nucleon mass.



\begin{thebibliography}{00}

\bibitem{snp}
For a recent collection of reviews on nuclear physics
with strangeness see the Special Issue on {\em Recent Advances
in Strangeness Nuclear Physics},
Nucl. Phys. {\bf A 804} (2008) 1.

\bibitem{ra98}
E. Oset and A. Ramos, Prog. Part. Nucl. Phys. {\bf 41} (1998) 191.

\bibitem{al02}
W. M. Alberico and G. Garbarino, Phys. Rep. {\bf 369} (2002) 1; in
{\em Hadron Physics}, IOS Press, Amsterdam, 2005, p.~125. Edited
by T. Bressani, A. Filippi and U. Wiedner. Proceedings of the
International School of Physics ``Enrico Fermi", Course CLVIII,
Varenna (Italy), June 22 -- July 2, 2004
[nucl-th/0410059]. For an update of these reviews see
Ref~.\cite{Ch08}.

\bibitem{Ch08}
C. Chumillas, G. Garbarino, A. Parre\~no and A. Ramos,
Nucl. Phys. {\bf A 804} (2008) 162.

\bibitem{Pa07}
A. Parre\~no, Lect. Notes Phys. {\bf 724} (2007) 141.

\bibitem{Ou05}
H. Outa, in {\em Hadron Physics} (IOS Press, Amsterdam, 2005) p.~219.
Edited by T. Bressani, A. Filippi and U. Wiedner.
Proceedings of the International School of Physics
``Enrico Fermi", Course CLVIII, Varenna (Italy), June 22 -- July 2, 2004.

\bibitem {mesonic-th}
J. Nieves and E. Oset, Phys.Rev. {\bf C 47} (1993) 1478;
T. Motoba and K. Itonaga, Prog. Theor. Phys. Suppl. {\bf 117} (1994) 477;
T. Motoba, K. Itonaga and H. Bando, Nucl. Phys. {\bf A 489} (1988) 683.

\bibitem{gal09}
A. Gal, Nucl. Phys. {\bf A 828} (2009) 72.

\bibitem {ra97}
A. Ramos, M. J. Vicente-Vacas and E. Oset, Phys. Rev. {\bf C 55}
(1997) 735; {\bf 66} (2002) 039903(E).

\bibitem{ga03}
G. Garbarino, A. Parre\~no and A. Ramos, Phys. Rev. Lett. {\bf 91}
(2003) 112501.

\bibitem{ga04}
G. Garbarino, A. Parre\~no and A. Ramos, Phys. Rev. {\bf C 69}
(2004) 054603.

\bibitem {ba03}
E. Bauer and F. Krmpoti\'c, Nucl. Phys. {\bf A 717} (2003) 217.

\bibitem{ba04}
E. Bauer and F. Krmpoti\'c, Nucl. Phys. {\bf A 739} (2004) 109.

\bibitem{KEK}
B. H. Kang et al.,  Phys. Rev. Lett. {\bf 96} (2006) 062301.

\bibitem{KEK2}
M. J. Kim et al., Phys. Lett. {\bf B 641} (2006) 28.

\bibitem{Ba06}
E. Bauer, G. Garbarino, A. Parre\~no and A. Ramos,
e-Print nucl-th/0602066.

\bibitem {os85}
E. Oset and L. L. Salcedo, Nucl. Phys. {\bf A 443} (1985) 704.

\bibitem{ba09}
E. Bauer, Nucl. Phys. {\bf A 818} (2009) 174.

\bibitem{ba09b}
E. Bauer and G. Garbarino,
Nucl. Phys. {\bf A 828} (2009) 29.

\bibitem{Ji01}
D. Jido, E. Oset and J. E. Palomar, Nucl. Phys. {\bf A 694} (2001) 525.

\bibitem{ba07b}
E. Bauer, Nucl. Phys. {\bf A 796} (2007) 11.

\bibitem {pa97}
A. Parre\~no, A. Ramos and C. Bennhold, Phys. Rev. {\bf C 56}
(1997) 339; \\
A. Parre\~{n}o and A. Ramos, Phys. Rev. {\bf C 65} (2002) 015204.

\bibitem {st99}
V. G. J. Stoks and Th. A. Rijken, Phys. Rev. {\bf C 59} (1999)
3009; Th. A. Rijken, V. G. J. Stoks and Y. Yamamoto, {\it ibid.}
{59} (1999) 21.

\bibitem {ma87}
R. Machleidt, K. Holinde and Ch. Elster; Phys. Rep. {\bf 149}
(1987) 1.

\bibitem {br96}
M. B. Barbaro, A. De Pace, T. W. Donnelly and A. Molinari, Nucl.
Phys. {\bf A 596} (1996) 553.

\bibitem{Ki09}
M. Kim et al., Phys. Rev. Lett. {\bf 103}, 182502 (2009).

\bibitem{Ag09}
M. Agnello et al., e-Print: arXiv:0910.4939 [nucl-ex].

\bibitem{Cosy}
W. Cassing et al., Eur. Phys. J. {\bf A 16}, 549 (2003).

\bibitem{Sa05}
Y. Sato et al., Phys. Rev. {\bf C 71}, 025203 (2005).

\bibitem{Ou08}
H. Outa, contributed talk at the PANIC08 conference,
Eilat (Israel), November 9--14, 2008;
H. Outa et al., Nucl. Phys. {\bf A 754} (2005) 157c.

\bibitem{FI}
M. Agnello et al., Phys. Lett. {\bf B 622} (2005) 35;
Nucl. Phys. {\bf A 804} (2008) 151.

\bibitem{jparc}
T. Nagae, Nucl. Phys. {\bf A 754} (2005) 443c;
Nucl.Phys. {\bf A 805} (2008) 486c.

\bibitem{jparc-exp}
S. Ajimura et al., {\em Exclusive study on the $\Lambda N$ weak interaction in $A=4$
$\Lambda$ hypernuclei}, Letter of intent for an experiment (E22) at J--PARC (2007);
H. Bhang et al., {\em Coincidence measurement of the weak decay of
$^{12}_\Lambda$C and the three--body weak interaction process},
Letter of intent for an experiment (E18) at J--PARC (2006).

\bibitem{GSI}
T. Fukuda et al., Nucl. Phys. {\bf A 790} (2007) 161c.

\end{thebibliography}
\end{document}